\documentclass[%twocolumn,           % Format : 
onecolumn, %twocolumn
               showpacs,            % Pacs : showpacs, noshowpacs
               nopreprintnumbers,     % Preprint: preprintnumbers,
               aps,                 % Society: ...
               prd,          	    % Journal Style : pra, prb, prc, prd, pre,
               letterpaper,             % Size : a4paper, ...
              groupeaddress,      % Affiliation (Title) : groupedaddress,
               nofootinbib,         % Footnote: footinbib, nofootinbib
               tightenlines,        % Remove additional spaces in a line
               showkeys, 11pt,
               notitlepage
               ]{revtex4-1}
               
\usepackage{bm}% bold math
\usepackage{amsmath}
\usepackage{amsfonts,amssymb}
\usepackage{mathrsfs}
\usepackage{mathtools}
\usepackage{hyperref}
\usepackage{soul}
\usepackage{enumerate}
\usepackage{ulem}
\DeclareFontFamily{U}{BOONDOX-cal}{\skewchar\font=45 }
\DeclareFontShape{U}{BOONDOX-cal}{m}{n}{
  <-> s*[1.05] BOONDOX-r-cal}{}
\DeclareFontShape{U}{BOONDOX-cal}{b}{n}{
  <-> s*[1.05] BOONDOX-b-cal}{}
\DeclareMathAlphabet{\mathcalboondox}{U}{BOONDOX-cal}{m}{n}
\SetMathAlphabet{\mathcalboondox}{bold}{U}{BOONDOX-cal}{b}{n}
\DeclareMathAlphabet{\mathbcalboondox}{U}{BOONDOX-cal}{b}{n}
\newcommand{\mcb}[1]{\mathcalboondox{#1}}
\newcommand{\mc}[1]{\mathcal{#1}}

\begin{document}

\title{General perturbations in D+1 standard and brane cosmology revisited}

\author{Zo\'e Delf\'in$^1$}
%\email{delfibranas@gmail.com}

\author{Rub\'en Cordero$^1$}
%\email{cordero@esfm.ipn.mx}

\author{Tonatiuh Matos$^2$}
%\email{tmatos@fisica.cinvestav.mx}

\author{Miguel A. Garc\'ia-Aspeitia$^{3}$}
%\email{angel.garcia@ibero.mx}

\affiliation{$^1$Departamento de F\'isica, Escuela Superior de F\'iısica
y Matem\'aticas del Instituto Polit\'ecnico Nacional,
Unidad Adolfo L\'opez Mateos, Edificio 9, 07738, CDMX, M\'exico\\
delfibranas@gmail.com and cordero@esfm.ipn.mx}
\affiliation{$^2$Departamento de F\'isica, Centro de Investigaci\'on y de Estudios Avanzados del IPN,
A.P. 14-740, 07000 CDMX, M\'exico.\\
tonatiuh.matos@cinvestav.mx}
\affiliation{$^3$Depto. de Física y Matemáticas, Universidad Iberoamericana Ciudad de M\'exico, Prolongación Paseo \\ de la Reforma 880, 01219, CDMX, México.\\
angel.garcia@ibero.mx}

\begin{abstract}
In the present work, we generalize the theory of perturbations in a $D+1$ dimensional space-time with cosmological constant, studying scalar, vector and tensor perturbations, as well as its structure in Newtonian and Synchronous gauge. We also show the theory of perturbations in the context of brane cosmology, where branes are embedded in a set of D-spatial dimensions, a temporal dimension, and an additional spatial dimension. In both, standard and brane cosmology an unperturbed spacetime is provided with a Friedmann-Lemaitre-Robertson-Walker metric and arbitrary sectional curvature, the matter content has the shape of a perfect fluid. In addition, we consider the arbitrary sectional curvature, obtaining the respective equations in the Newtonian and Synchronous gauge. We highlight that the results presented in this article can be used to treat brane cosmology with two concentric branes or tackle the $\sigma_8$ tension with a braneworld approach. Finally, as an example of the utility of all the technology presented in this paper, we show an application to the energy flux and the repercussions in the framework of the brane-worlds.
\end{abstract}

\maketitle

%%%%%%%%%%%%%%%%%%%%%%%%%%%%%%%%%%%%
\section{Introduction}
%%%%%%%%%%%%%%%%%%%%%%%%%%%%%%%%%%%%

Braneworlds have been extensively studied in the literature to solve problems in scenarios associated with particle physics and cosmology. In cosmology, the main objectives are to address the problem of the observed acceleration of the Universe \cite{Riess:1998,Perlmutter:1999} and the nature of dark matter, which have not yet been resolved. For now, the best candidate to understand the puzzle of accelerating expansion is the well-known cosmological constant (CC). For dark matter, the best candidates for this missing mass are supposed to be dust or an ultralight scalar field \cite{Matos:1998vk,Matos:2000ss}. However, we do not have any detection of any of them, and the CC is plagued of unsolved problems like the fine tuning and coincidence problems (see \cite{Carroll:2000} for details), seeing ourselves in need to change the paradigm and explore other alternatives for the Universe acceleration \cite{Motta:2021hvl} and the dark matter.

The idea of incorporate extra dimension was born due to the hierarchy problem associated with the weakness of gravity in comparison with the other forces. From an heuristic point of view, the addition of extra dimensions generates corrections in the gravitational potential that scales as $\sim r^{-(d+1)}$ where $d$ is the number of dimensions. In this case, several experiments contribute with stringent values for the existence of extra dimensions \cite{Long:2003dx}. 
The Randall-Sundrum (RS) models incorporates one or two cosmological branes into a five dimensional bulk with anti-de Sitter geometry, with the aim of alleviate the hierarchy problem, in this case, the RS model with only one brane is more successful due to its simplicity \cite{Randall:1999ee,Randall:1999vf}. Thus, under this background, it is possible to construct the covariant approach in where a modified Einstein equation rise with extra terms that provides with extra richness to the behavior of gravity in our brane. We mention also the Arkani-Hamed-Dimopoulos-Dvali (ADD) model \cite{Arkani-Hamed:1998jmv,Antoniadis:1998ig,Akama:1982jy} in where the idea of a large volume for the compact extra dimensions would generates changes in the fundamental Planck scale. The Dvali-Gabadadze-Porrati (DGP) \cite{Dvali:2000hr} is another brane model in where the transition to a five dimensional space time mimics the observed Universe acceleration. However, when the equations are transferred to a cosmological context and put into the test using cosmological data samples, the DGP model present conflict with observables \cite{Fang:2008kc}.  

The covariant form of the Einstein equation modifies the standard model of cosmology, existing new fields like the well known {\it dark radiation} that comes from the projection of the Weyl term in five dimensions. In the same fashion as happens in standard cosmology, in braneworlds linear perturbations of the modified field equations are used in order to study the acoustic peaks of the cosmic microwave background radiation (CMB) noticing changes mainly in the first acoustic peak  \cite{Koyama:2003be}. From the astrophysics and cosmological point of view, the gravitational waves generated in the braneworld context also have been studied and in particular in the inflationary scenario \cite{Frolov:2002qm}. Recent studies shown that braneworld cosmologies based in RS context present difficulties due to the tension observed in the constraints of the free parameters when it is exposed to cosmological data samples \cite{Garcia-Aspeitia:2016kak}. However the braneworlds based in RS model but with a variable brane tension are successful in alleviate the tension with observables and additionally, the brane tension acts like the causative of the current Universe acceleration \cite{Garcia-Aspeitia:2018fvw}. We encourage to the reader the revision of \cite{Maartens:2010ar} in where it is presented several brane world models.

Linear perturbations are required to study effects related with the growth of structure, CMB, inflation, among others. In particular, for the extra dimensional case, perturbations are extensively studied in several references (see for instance \cite{Maartens:2000fg,vandeBruck:2000ju,Koyama:2000cc,Langlois:2000ph,Deruelle:2000yj,Langlois:2000iu,Hwang:2001zt,Hwang:2002fp,Giudice:2002vh,Casali:2004dx,Yoshiguchi:2004nm,Parameswaran:2009bt,Dotti:2005sq,Gleiser:2005ra,Beroiz:2007gp,Konoplya:2008ix} and references therein), studying the consequences of brane tension, the dilaton dynamics, inflation, the tachyon condensation, among others. In order to mention some studies is for example the case of Ref. \cite{Maartens:2000fg} in where it is described the case of perturbations and in particular, pointing out the vector perturbations which its dilution is not in the same way as in the standard case. Other is for example, in Ref. \cite{Koyama:2000cc}, in where additionally to the perturbation study, they remark the main differences with the standard 4D knowledge.

These kind of studies are the cornerstone to understand the effects of extra dimensions in the Universe evolution in order to compare with recent observations provided by different telescopes. Additionally to this, linear perturbations provide with restrictive constraints of the brane tension effects, no local terms and quadratic terms, which are characteristic of theories such as those shown in this investigation.
Thus, in the present work, we are interested in developing a formalism where the branes when interacting generate perturbations that later give rise to the formation of structures or gravitational waves. The branes studied in this paper can be constituted of arbitrary matter (for example, scalar fields for dark matter or spin 1 fields for the interactions of the standard model of particles).
We present a revision, beginning with perturbations with a generalization of $D+1$ space-time, showing the scalar, vector and tensor modes together with its expressions written in different gauges. In addition, we show the brane perturbations embedded in a $D+1$ dimensional bulk assuming a Friedmann-Lemaitre-Robertson-Walker (FLRW) metric geometry for the brane, where the Newtonian and Synchronous gauges for perturbations are also shown. The geometrical structure of the configuration studied in this paper is described as follows: We chose a generic geometrical structure of the bulk $(D+1)+1$ ($D+1$ space-time plus $1$ extra spatial coordinate) always under the restriction of homogeneity and isotropy in the spatial part. Furthermore, we consider an embedded brane $D+1$ ($D$ spatial plus $1$ temporal coordinate) restricted also to homogeneity and isotropy. We consider that the bulk has $t$-time while for the brane it has a proper time $\tau$. 
This mentioned structure, could be easily applied to cosmological models and in particular for the $\sigma_8$ tension in which exist discrepancies among CMB Planck results and weak gravitational lensing measurements (see \cite{Planck:2018,KiDS:2020suj}). Traditionally, the study of linear perturbations in extra dimensions has a particular geometry for the bulk and the branes. Thus, this generalization allow us to study a diversity of models including cosmology, black holes and especially gravitational waves, among others. Finally, notice that our expression can help with a more efficient computational analysis when it is studied the cosmic background radiation, structure formation and inflation.

The outline of the paper is in the following form: in Sec. \ref{sec:PertCosmoStandar} we revisit the perturbations in $D+1$ dimensions, adding new information lacked in other papers, showing the scalar, vector and tensor modes. In Sec. \ref{sec:Gauges} we present the different gauges centering our attention in the Newtonian and Synchronous gauges. In Sec. \ref{sec:PertCosmoBrane} we develop a study of perturbations on a D+1-dimensional brane embedded in a (D+1)+1-dimensional bulk, in similarity to that realized in standard cosmology. Sec. \ref{sec:GaugesBrane} is devoted to present the Newtonian and Synchronous gauges for a brane. In all sections, we consider that the spatial part of the metrics is of $D$ dimension and with arbitrary sectional curvature. The linear perturbations presented in this paper has application in models like these associated with two concentric branes and in the analysis of the $\sigma_8$ tension associated with linear perturbations. Finally in Sec. \ref{Sec:CO} we present the discussion and conclusions. We henceforth use units in which $\hbar=c=k_{B}=1$ (natural units).

%%%%%%%%%%%%%%%%%%%%%%%%%%%%%%%%%%%%
\section{Perturbations in $D+1$ dimensions} \label{sec:PertCosmoStandar}
%%%%%%%%%%%%%%%%%%%%%%%%%%%%%%%%%%%%

We initiate our analysis with the line element for this space-time which reads as
\begin{eqnarray}
ds^{2}&=&\gamma_{\tau\tau}d\tau^{2}+\gamma_{ij}dx^{i}dx^{j}=-N^{2}d\tau^{2}+a^{2}\tilde{\gamma}_{ij}dx^{i}dx^{j}. \label{eq:Metrica}
\end{eqnarray} 
Considering small perturbations around the metric $\gamma_{\mu\nu}$ in $D+1$ dimensions \cite{weinbergcosmology,dodelsoncosmology}, we have  $
\bar{\gamma}_{\mu\nu}=\gamma_{\mu\nu}+\eta_{\mu\nu}$, where $\gamma_{\mu\nu}$ is the unperturbed metric tensor and $\eta_{\mu\nu}\ll1$ is the perturbation which is symmetric under index commutations as $\eta_{\mu\nu}=\eta_{\nu\mu}$. All perturbed quantities will be denoted with a bar, in the form
$\bar{\gamma}^{\mu\nu}=\gamma^{\mu\nu}+\eta^{\mu\nu}$, $\bar{\gamma}^{\mu}_{\ \nu}=\gamma^{\mu}_{\ \nu}$. Considering the first order perturbation under the Friedmann-Lemaitre-Robertson-Walker (FLRW) $D+1$ metric, with arbitrary sectional curvature denoted by $K$, we have that the metric tensor can be written as follows
$\gamma_{\tau\tau}=-\textsl{N}^{2}$ , $\gamma_{i\tau}=\gamma_{\tau i}=0$, $\gamma_{ij}=a^{2} \tilde{\gamma}_{ij}$, where $a\equiv a(\tau)$ is the scale factor and $N=a$ (for the conformal time) and $N=1$ (for the cosmic time). We use the tilde notation to indicate that the object in question is purely spatial. Perturbations fulfill the following relations
\begin{eqnarray}
&\eta^{\mu\nu}=-\gamma^{\mu\rho}\gamma^{\nu\sigma}\eta_{\rho\sigma}, \quad \eta_{\mu\nu}=-\gamma_{\mu\rho}\gamma_{\nu\sigma}\eta^{\rho\sigma}, \quad \gamma^{\mu\sigma}\eta_{\sigma\nu}=-\eta^{\mu\sigma}\gamma_{\sigma\nu}&.
\label{eq:RelacionesPerturbacion}
\end{eqnarray}
In order to construct the perturbed $D+1$ Einstein field equation, we proceed to find the perturbed Christoffel symbols through the formula
\begin{eqnarray}
\delta\Gamma^{\mu}_{\nu\lambda}=\frac{1}{2}\gamma^{\mu\rho}\left[-2 \eta_{\rho\sigma}\Gamma^{\sigma}_{\nu\lambda}+\partial_{\lambda}\eta_{\rho\nu}+\partial_{\nu}\eta_{\rho\lambda}-\partial_{\rho}\eta_{\nu\lambda} \right],
\label{eq:PerturbacionChristoffel}
\end{eqnarray}
here $\delta$ indicates perturbation. Therefore the non-null perturbed symbols are
\begin{eqnarray}
\delta\Gamma^{\tau }_{\tau \tau }&=&\frac{1}{2\textsl{N}^{2}}\left(2\frac{\dot{\textsl{N}}}{\textsl{N}}\eta_{\tau \tau }-\dot{\eta}_{\tau \tau }\right), \nonumber \\
\delta\Gamma^{\tau }_{\tau i}&=&\frac{1}{2\textsl{N}^{2}}\left(2\frac{\dot{a}}{a}\eta_{i\tau }-\partial_{i}\eta_{\tau \tau }\right), \nonumber \\
\delta\Gamma^{\tau }_{ij}&=&\frac{1}{2\textsl{N}^{2}}\left(2\frac{a\dot{a}}{\textsl{N}^{2}}\tilde{\gamma}_{ij}\eta_{\tau \tau }-\nabla_{j}\eta_{i\tau }-\nabla_{i}\eta_{j\tau }+\dot{\eta}_{ij} \right), \ \ \ \nonumber \\
\delta\Gamma^{i}_{\tau \tau }&=&\frac{1}{2a^{2}}\tilde{\gamma}^{im} \left(-2\frac{\dot{\textsl{N}}}{\textsl{N}}\eta_{m\tau }+2\dot{\eta}_{m\tau }-\partial_{m}\eta_{\tau \tau } \right), \nonumber \\
\delta\Gamma^{i}_{j\tau }&=&\frac{1}{2a^{2}}\tilde{\gamma}^{im}\left( -2\frac{\dot{a}}{a}\eta_{mj}+\dot{\eta}_{mj}+\nabla_{j}\eta_{m\tau }-\nabla_{m}\eta_{j\tau } \right), \nonumber \\
\delta\Gamma^{i}_{jk}&=&\frac{1}{2a^{2}}\tilde{\gamma}^{im} \Bigl( -2\frac{a\dot{a}}{\textsl{N}^{2}}\tilde{\gamma}_{jk}\eta_{m\tau }+\nabla_{k}\eta_{mj}+\nabla_{j}\eta_{mk}-\nabla_{m}\eta_{jk} \Bigr), 
\label{eq:ChristoffelRWPert}
\end{eqnarray}
where $\dot{\eta}_{\mu\nu}$ and $\nabla_{\lambda}{\eta}_{\mu\nu}$ are the temporal and covariant derivatives of $\eta_{\mu\nu}$, respectively. Moreover, the perturbed Ricci curvature tensor can be calculated through the formula
\begin{eqnarray}
\delta R_{\mu\kappa}&=&\partial_{\kappa}\delta\Gamma^{\lambda}_{\lambda\mu}-\partial_{\lambda}\delta\Gamma^{\lambda}_{\mu\kappa}+\Gamma^{\lambda}_{\mu\sigma}\delta\Gamma^{\sigma}_{\kappa\lambda}+\Gamma^{\sigma}_{\kappa\lambda}\delta\Gamma^{\lambda}_{\mu\sigma}-\Gamma^{\lambda}_{\mu\kappa}\delta\Gamma^{\sigma}_{\sigma\lambda}-\delta\Gamma^{\lambda}_{\mu\kappa}\Gamma^{\sigma}_{\sigma\lambda},
\end{eqnarray}
thus
\begin{eqnarray}
\delta R_{\tau \tau }&=&\frac{1}{2a^{2}}\tilde{\gamma}^{ml}\nabla_{m}\nabla_{l}\eta_{\tau \tau }+\frac{D\dot{a}}{2aN^{2}}\left(\dot{\eta}_{\tau \tau }-2\frac{\dot{N}}{N}\eta_{\tau \tau }\right)\nonumber \\
& &+\frac{1}{2a^{2}}\tilde{\gamma}^{lm}\left[\ddot{\eta}_{ml}-\left(\frac{2\dot{a}}{a}+\frac{\dot{N}}{N}\right)\dot{\eta}_{ml}+2\left(\frac{\dot{a}^{2}}{a^{2}}+\frac{\dot{a}\dot{N}}{aN}-\frac{\ddot{a}}{a} \right)\eta_{ml}\right]\nonumber \\
& &-\frac{1}{a^{2}}\tilde{\gamma}^{ml}\nabla_{m}\left(\dot{\eta}_{l\tau }-\frac{\dot{N}}{N}\eta_{l\tau }\right),  \\ \nonumber \\ \nonumber \\
\delta R_{\tau j}&=&\frac{\left(D-1\right)\dot{a}}{2aN^{2}}\nabla_{j}\eta_{\tau \tau }+\frac{1}{2a^{2}}\tilde{\gamma}^{ml}\left(\nabla_{m}\nabla_{l}\eta_{j\tau }-\nabla_{j}\nabla_{l}\eta_{m\tau }\right) \nonumber \\
& &-\left[\frac{\ddot{a}}{aN^{2}}+\left( D-1\right)\frac{\dot{a}^{2}}{a^{2}N^{2}}-\frac{\dot{a}\dot{N}}{aN^{3}}\right]\eta_{j\tau }+\frac{1}{2}\frac{\partial}{\partial t}\left[\frac{1}{a^{2}}\gamma^{ml}\left(\nabla_{j}\eta_{ml}-\nabla_{l}\eta_{mj} \right)   \right], \nonumber \\
\end{eqnarray}

\begin{eqnarray}
\delta R_{jk}&=&-\frac{1}{2N^{2}}\nabla_{j}\nabla_{k}\eta_{\tau \tau }-\left[ \left(D-1\right)\frac{\dot{a}^{2}}{N^{4}}+\frac{a\ddot{a}}{N^{4}}-2\frac{a\dot{a}\dot{N}}{N^{5}} \right] \tilde{\gamma}_{jk}\eta_{\tau \tau }-\frac{1}{2}\frac{a\dot{a}}{N^{4}}\tilde{\gamma}_{jk}\dot{\eta}_{\tau \tau } \nonumber \\
& &+\frac{1}{2a^{2}}\left[ \tilde{\gamma}^{nm}\nabla_{n}\nabla_{m}\eta_{jk}-\nabla_{n}\nabla_{j}\left(\tilde{\gamma}^{nm}\eta_{mk}\right)\right. \nonumber \\
& &\left.-\nabla_{n}\nabla_{k}\left(\tilde{\gamma}^{nm}\eta_{mj}\right)+\nabla_{j}\nabla_{k}\left(\tilde{\gamma}^{nm}\eta_{nm}\right) \right]  \nonumber\\
& &-\frac{1}{2N^{2}}\ddot{\eta}_{jk}+\left[ \left(4-D\right)\frac{\dot{a}}{2aN^{2}}+\frac{\dot{N}}{2N^{3}}\right]\dot{\eta}_{jk}-\frac{\dot{a}}{2aN^{2}}\tilde{\gamma}_{jk}\tilde{\gamma}^{nm}\dot{\eta}_{nm} \nonumber \\
& &+\frac{\dot{a}^{2}}{a^{2}N^{2}}\left( -2\eta_{jk}+\tilde{\gamma}_{jk}\tilde{\gamma}^{nm}\eta_{nm}  \right)+\frac{\dot{a}}{aN^{2}}\tilde{\gamma}_{jk}\tilde{\gamma}^{nm}\nabla_{n}\eta_{m\tau } \nonumber \\
& &+\frac{1}{2N^{2}}\left( \nabla_{j}\dot{\eta}_{k\tau }+\nabla_{k}\dot{\eta}_{j\tau } \right)+\frac{1}{2N^{2}}\left[\left( D-2 \right)\frac{\dot{a}}{a}-\frac{\dot{N}}{N}\right]\left( \nabla_{j}\eta_{k\tau }+\nabla_{k}\eta_{j\tau } \right). \nonumber \\
\end{eqnarray}
Until now, we have developed the geometric side of the perturbed part of Einstein's equation. To continue with the matter side, we will first work with the unperturbed equations. 

Einstein equation can take the following form
\begin{eqnarray}
R_{\mu\nu}=\kappa S_{\mu\nu},
\label{eq:EcEinstein}
\end{eqnarray}
where $\kappa=-8\pi G$, $G$ is the Newton gravitational constant and $S_{\mu\nu}$ is the source tensor and can be written in terms of the energy-momentum tensor and $\Lambda$ as the cosmological constant,
\begin{eqnarray}
S_{\mu\nu}\equiv T_{\mu\nu}-\frac{1}{D-1}\left(\gamma_{\mu\nu}T^{\lambda}_{\ \lambda}-\frac{2}{\kappa}\Lambda \gamma_{\mu\nu}\right).
\end{eqnarray}
We consider that the unperturbed energy-momentum tensor must be in the form of a perfect fluid as
\begin{eqnarray}
T_{\mu\nu}=P\gamma_{\mu\nu}+\left( P+\rho\right)u_{\mu}u_{\nu},
\end{eqnarray}
where $\rho$ is the energy density, $P$ is the pressure and the D+1 velocity components are $u_{\tau }=-N$ and $u_{i}=0$ in a comoving system, then, the components of the energy-momentum tensor are
\begin{eqnarray}
&T_{\tau \tau }=N^{2}\rho, \quad T_{\tau i}=T_{i\tau }=0, \quad T_{ij}=T_{ji}=a^{2}P\tilde{\gamma}_{ij},& \nonumber \\[5pt]
&T^{\tau }_{\ \tau }=-\rho, \quad T^{i}_{\ j}=P\delta^{i}_{j}, \quad T^{k}_{\ k}=DP, \quad T^{\lambda}_{\ \lambda}=-\rho+DP&,
\label{eq:ComponentesTensorFluidoPerfecto}
\end{eqnarray}
therefore the components of the source tensor are
\begin{eqnarray}
S_{ij}=\frac{1}{D-1}\left(\rho-P+\frac{2}{\kappa}\Lambda\right)a^{2}\tilde{\gamma}_{ij}, \quad S_{i\tau }=0, \nonumber \\[5pt] S_{\tau \tau }=\frac{1}{D-1}\left[\left(D-2\right)\rho+DP-\frac{2}{\kappa}\Lambda\right]N^{2},
\label{eq:TensorFuente}
\end{eqnarray}
and the components of the Ricci tensor are
\begin{eqnarray}
R_{ij}=\left[\left(1-D\right)\left(K+\frac{\dot{a}^{2}}{N^{2}}\right)-\frac{a\ddot{a}}{N^{2}}\right] \tilde{\gamma}_{ij}, \quad R_{i\tau }=0, \quad R_{\tau \tau }=D\left(-\frac{\dot{a}\dot{N}}{aN}+\frac{\ddot{a}}{a}\right).
\label{eq:TensorRicci}
\end{eqnarray}
By substituting \eqref{eq:TensorFuente} and \eqref{eq:TensorRicci} into Eq. \eqref{eq:EcEinstein} we obtain the components of the Einstein equation
\begin{eqnarray}
&\left(1-D\right)\left(K+\frac{\dot{a}^{2}}{N^{2}}\right)-\frac{a\ddot{a}}{N^{2}}=\frac{\kappa}{D-1}\left(\rho-P+\frac{2}{\kappa}\Lambda\right)a^{2}, \nonumber \\[5pt]
&D\left(-\frac{\dot{a}\dot{N}}{aN}+\frac{\ddot{a}}{a}\right)=\frac{\kappa}{D-1}\left[\left(D-2\right)\rho+DP-\frac{2}{\kappa}\Lambda\right]N^{2}, &
\end{eqnarray}
thus we obtain the Friedmann equations in $D+1$ dimensions as
\begin{eqnarray}
\rho=\frac{D}{2\kappa}\left[\left(\frac{1-D}{a^{2}}\right)\left(K+\frac{\dot{a}^{2}}{N^{2}}\right)-\frac{\dot{a}\dot{N}}{aN^{3}}-\frac{2\Lambda}{D}\right],
\label{eq:Densidad}
\end{eqnarray}

\begin{eqnarray}
P=\frac{1}{2\kappa}\left[\frac{\left(D-2\right)\left(D-1\right)}{a^{2}}\left(K+\frac{\dot{a}^{2}}{N^{2}}\right)-D\frac{\dot{a}\dot{N}}{aN^{3}}+2\left(D-1\right)\frac{\ddot{a}}{aN^{2}}+2\Lambda\right].
\label{eq:Presion}
\end{eqnarray}
We continue developing the perturbation of the source tensor similarly as the previous objects in the form
\begin{eqnarray}
\delta S_{\mu\nu}=\delta T_{\mu\nu}-\frac{1}{D-1}\left[\gamma_{\mu\nu}\delta T^{\lambda}_{\ \lambda}+\left( T^{\lambda}_{\ \lambda}-\frac{2}{\kappa}\Lambda\right)\eta_{\mu\nu}\right],
\label{eq:PertTensorFuente}
\end{eqnarray}
and substituting \eqref{eq:Densidad} and \eqref{eq:Presion} in \eqref{eq:PertTensorFuente} we have
{\normalsize
\begin{eqnarray}
\delta S_{\tau \tau }&=&\delta T_{\tau \tau }+\frac{1}{D-1}N^{2}\delta T^{\lambda}_{\ \lambda} \nonumber \\[5pt]
&&-\frac{D}{2\kappa}\left[\frac{\left(D-1\right)}{a^{2}}\left(K+\frac{\dot{a}^{2}}{N^{2}}\right)-\frac{\dot{a}\dot{N}}{aN^{3}}+2\frac{\ddot{a}}{aN^{2}}-\frac{4}{D\left(D-1\right)}\Lambda\right]\eta_{\tau \tau },
\end{eqnarray}

%\\[3pt]

\begin{eqnarray}
\delta S_{\tau j}&=&\delta S_{j\tau }=\delta T_{j\tau } \nonumber \\[5pt]
&&-\frac{D}{2\kappa}\left[\frac{\left(D-1\right)}{a^{2}}\left(K+\frac{\dot{a}^{2}}{N^{2}}\right)-\frac{\dot{a}\dot{N}}{aN^{3}}+2\frac{\ddot{a}}{aN^{2}}-\frac{4}{D\left(D-1\right)}\Lambda\right]\eta_{j\tau },
\end{eqnarray}

%\\[3pt]

\begin{eqnarray}
\delta S_{ij}&=&\delta T_{ij}-\frac{1}{D-1}a^{2}\tilde{\gamma}_{ij}\delta T^{\lambda}_{\ \lambda}\nonumber \\[5pt]
&&-\frac{D}{2\kappa}\left[\frac{\left(D-1\right)}{a^{2}}\left(K+\frac{\dot{a}^{2}}{N^{2}}\right)-\frac{\dot{a}\dot{N}}{aN^{3}}+2\frac{\ddot{a}}{aN^{2}}-\frac{4}{D\left(D-1\right)}\Lambda\right]\eta_{ij}.
\end{eqnarray}
}

Putting together all the elements to present the perturbed version of Einstein equations give us
{\normalsize
\begin{eqnarray}
&&\kappa\left(\delta T_{\tau \tau }+\frac{1}{D-1}N^{2}\delta T^{\lambda}_{\ \lambda}\right)= \nonumber \\[5pt]
&&\frac{D}{2}\left[\frac{\left(D-1\right)}{a^{2}}\left(K+\frac{\dot{a}^{2}}{N^{2}}\right)-3\frac{\dot{a}\dot{N}}{aN^{3}}+2\frac{\ddot{a}}{aN^{2}}-\frac{4}{D\left(D-1\right)}\Lambda\right]\eta_{\tau \tau }\nonumber \\
&&+\frac{1}{2a^{2}}\tilde{\gamma}^{ml}\nabla_{m}\nabla_{l}\eta_{\tau \tau }+\frac{D\dot{a}}{2aN^{2}}\dot{\eta}_{\tau \tau }\nonumber \\
&&+\frac{1}{2a^{2}}\tilde{\gamma}^{lm}\left[\ddot{\eta}_{ml}-\left(\frac{2\dot{a}}{a}+\frac{\dot{N}}{N}\right)\dot{\eta}_{ml}+2\left(\frac{\dot{a}^{2}}{a^{2}}+\frac{\dot{a}\dot{N}}{aN}-\frac{\ddot{a}}{a} \right)\eta_{ml}\right]\nonumber \\
&&-\frac{1}{a^{2}}\tilde{\gamma}^{ml}\nabla_{m}\left(\dot{\eta}_{l\tau }-\frac{\dot{N}}{N}\eta_{l\tau }\right),
\label{eq:EcEinsteinScalar}
\end{eqnarray}
\\
\begin{eqnarray}
&&\kappa\delta T_{\tau j}=\frac{\left(D-1\right)\dot{a}}{2aN^{2}}\nabla_{j}\eta_{\tau \tau }+\frac{1}{2a^{2}}\tilde{\gamma}^{ml}\left(\nabla_{m}\nabla_{l}\eta_{j\tau }-\nabla_{j}\nabla_{l}\eta_{m\tau }\right) \nonumber \\
& &+\frac{1}{2}\left[\frac{D\left(D-1\right)}{a^{2}}K+\left(D-2\right)\left(D-1\right)\frac{\dot{a}^{2}}{a^{2}N^{2}}-\left(D-2\right)\frac{\dot{a}\dot{N}}{aN^{3}} \right. \nonumber \\
& &+\left.2\left(D-1\right)\frac{\ddot{a}}{aN^{2}}-\frac{4}{\left(D-1\right)}\Lambda\right]\eta_{j\tau }+\frac{1}{2}\frac{\partial}{\partial t}\left[\frac{1}{a^{2}}\gamma^{ml}\left(\nabla_{j}\eta_{ml}-\nabla_{l}\eta_{mj} \right)   \right],
\label{eq:EcEinsteinVector}
\end{eqnarray}
}

{\normalsize
\begin{eqnarray}
&&\kappa\left(\delta T_{jk}-\frac{1}{D-1}a^{2}\tilde{\gamma}_{jk}\delta T^{\lambda}_{\ \lambda}\right)=-\frac{1}{2N^{2}}\nabla_{j}\nabla_{k}\eta_{\tau \tau } \nonumber \\
&&-\left[ \left(D-1\right)\frac{\dot{a}^{2}}{N^{4}}+\frac{a\ddot{a}}{N^{4}}-2\frac{a\dot{a}\dot{N}}{N^{5}} \right] \tilde{\gamma}_{jk}\eta_{\tau \tau }-\frac{1}{2}\frac{a\dot{a}}{N^{4}}\tilde{\gamma}_{jk}\dot{\eta}_{\tau \tau } \nonumber \\
& &+\frac{1}{2a^{2}}\left[\tilde{\gamma}^{nm}\nabla_{n}\nabla_{m}\eta_{jk}-\nabla_{n}\nabla_{j}\left(\tilde{\gamma}^{nm}\eta_{mk}\right)-\nabla_{n}\nabla_{k}\left(\tilde{\gamma}^{nm}\eta_{mj}\right)+\nabla_{j}\nabla_{k}\left(\tilde{\gamma}^{nm}\eta_{nm}\right)\right]  \nonumber\\  
& &-\frac{1}{2N^{2}}\ddot{\eta}_{jk}+\left[ \left(4-D\right)\frac{\dot{a}}{2aN^{2}}+\frac{\dot{N}}{2N^{3}}\right]\dot{\eta}_{jk}-\frac{\dot{a}}{2aN^{2}}\tilde{\gamma}_{jk}\tilde{\gamma}^{nm}\dot{\eta}_{nm} \nonumber \\
& &+\frac{\dot{a}^{2}}{a^{2}N^{2}}\tilde{\gamma}_{jk}\tilde{\gamma}^{nm}\eta_{nm}+\frac{\dot{a}}{aN^{2}}\tilde{\gamma}_{jk}\tilde{\gamma}^{nm}\nabla_{n}\eta_{m\tau } \nonumber \\
& &+\frac{1}{2N^{2}}\left( \nabla_{j}\dot{\eta}_{k\tau }+\nabla_{k}\dot{\eta}_{j\tau } \right)+\frac{1}{2N^{2}}\left[\left( D-2 \right)\frac{\dot{a}}{a}-\frac{\dot{N}}{N}\right]\left( \nabla_{j}\eta_{k\tau }+\nabla_{k}\eta_{j\tau } \right)+ \nonumber \\
& &\frac{D}{2}\left[\left(D-1\right)\frac{K}{a^{2}}+\left(D-1-\frac{4}{D}\right)\frac{\dot{a}^{2}}{a^{2}N^{2}}-\frac{\dot{a}\dot{N}}{aN^{3}}+2\frac{\ddot{a}}{aN^{2}}-\frac{4}{D\left(D-1\right)}\Lambda\right]\eta_{jk},
\label{eq:EcEinsteinTensor}
\end{eqnarray}
}
The energy-momentum tensor is restricted to the conservation condition $T^{\sigma}_{\ \nu; \sigma}=0$. The respective perturbed version is given by
\begin{eqnarray}
\partial_{\sigma}\delta T^{\sigma}_{\ \nu}+\Gamma^{\sigma}_{\sigma\rho}\delta T^{\rho}_{\ \nu}+\delta\Gamma^{\sigma}_{\sigma\rho}T^{\rho}_{\ \nu}-\Gamma^{\rho}_{\sigma\nu}\delta T^{\sigma}_{\ \rho}-\delta\Gamma^{\rho}_{\sigma\nu}T^{\sigma}_{\ \rho}=0,
\end{eqnarray}
setting $\nu=j$ gives the equation of momentum conservation
\begin{eqnarray}
&&\partial_{\tau }\delta T^{\tau }_{\ j}+\nabla_{l}\delta T^{l}_{\ j}+\left[\frac{\dot{N}}{N}+\left(D-1\right)\frac{\dot{a}}{a}\right]\delta T^{\tau }_{\ j}-\frac{a\dot{a}}{N^{2}}\tilde{\gamma}_{lj}\delta T^{l}_{\ \tau }\nonumber\\&&+\frac{1}{2N^{2}}\left(2\frac{\dot{a}}{a}\eta_{j\tau }-\nabla_{j}\eta_{\tau \tau }\right)\left(\rho+P\right)=0, 
\label{eq:ConservacionMomento}
\end{eqnarray}
and setting $\nu=0$ gives the equation of energy conservation
\begin{eqnarray}
\partial_{\tau }\delta T^{\tau }_{\ \tau }+\nabla_{l}\delta T^{l}_{\ \tau }+D\frac{\dot{a}}{a}\delta T^{\tau }_{\ \tau }-\frac{\dot{a}}{a}\delta T^{l}_{\ l}+\frac{1}{2a^{2}}\tilde{\gamma}^{lm}\left(2\frac{\dot{a}}{a}\eta_{ml}-\dot{\eta}_{ml}\right)\left(\rho+P\right)=0.
\label{eq:ConservacionEnergia}
\end{eqnarray}
Perturbation of the metric can be decomposed into scalars, divergenceless vectors and divergenceless traceless symmetric tensors, which are not coupled to each other
\begin{eqnarray}
\eta_{\tau \tau }&=&-N^{2}E, \nonumber \\
\eta_{\tau i}&=&\eta_{i\tau }=aN\left[\nabla_{i}F+G_{i}\right], \nonumber \\
\eta_{ij}&=&a^{2}\left[ A\tilde{\gamma}_{ij}+\nabla_{i}\nabla_{j}B+\nabla_{j}C_{i}+\nabla_{i}C_{j}+D_{ij} \right],
\label{eq:Descomposicion}
\end{eqnarray}
where $A$, $B$, $C_{i}$, $D_{ij}$, $E$, $F$ y $G_{i}$ are functions of $\vec{x}$ and $t$, satisfying the conditions
\begin{eqnarray}
\varepsilon_{ijk}\tilde{\gamma}^{jl}\nabla_{l}\tilde{\gamma}^{km}\nabla_{m}F=0, \quad \varepsilon_{ijk}\tilde{\gamma}^{jl}\nabla_{l}\tilde{\gamma}^{km}\nabla_{m}B=0, \nonumber \\
\nabla_{i}G^{i}=0, \quad \nabla_{i}C^{i}=0, \quad  \nabla_{i}D^{i}_{\ j}=0,  \nonumber\\
D_{ij}=D_{ji}, \quad D^{i}_{\ i}=0.
\label{eq:RotarDiverSimet}
\end{eqnarray}  
Let us consider that the perturbed energy-momentum tensor is also a perfect fluid, then
\begin{eqnarray}
\delta T_{\mu\nu}&=&P \eta_{\mu\nu}+\left( \rho+P \right)u_{\mu}\delta u_{\nu}+\left( \rho+P \right)\delta u_{\mu}u_{\nu}+\left( \delta\rho+\delta P \right)u_{\mu}u_{\nu}\nonumber \\
&&+\gamma_{\mu\nu}\delta P,
\end{eqnarray}
for this, we have the $D+1$ perturbed velocity $\bar{u}_{\mu}=u_{\mu}+\delta u_{\mu}$, which must comply with the normalization conditions $\gamma^{\mu\nu}u_{\mu}u_{\nu}=-1$, so we have the following relation
\begin{eqnarray}
\gamma^{\mu\nu}\delta u_{\mu}=-\frac{1}{2}\eta^{\mu\nu}u_{\mu} \quad {\rm or} \quad \delta u^{\mu}=\gamma^{\mu\sigma}\delta u_{\sigma}-\gamma^{\mu\lambda}\gamma^{\eta\varsigma}\eta^{\lambda\varsigma}u_{\eta},
\end{eqnarray}
then we have $\delta u_{\tau }=\eta_{\tau \tau }/2N$.\\

In general, the perturbation of velocity $\delta u_{i}$ can be decomposed into the gradient of a scalar potential $\delta u$ and a divergenceless vector $\delta u^{V}_{i}$. For the perturbation of the energy-momentum tensor we can add the terms $\nabla_{i} \nabla_{j} \pi^{S}$, $\nabla_{i} \pi^{V}_{j}\,+\,\nabla_{j} \pi^{V}_{i}$, and $\pi^{T}_{ij}$, which represent dissipative corrections to the inertia tensor and the following conditions are satisfied
\begin{equation}
\nabla_{i} \pi^{iV}=\nabla_{i} \delta u^{iV}=0, \quad \nabla_{i} \pi^{iT}_{\ j}=0, \quad \pi^{iT}_{\ i}=0,
\end{equation}
with these corrections, the perturbation of the energy-momentum tensor for a perfect fluid takes the following form
\begin{eqnarray}
&\delta T_{\tau \tau }=-\rho \eta_{\tau \tau }+N^{2}\delta\rho, \quad \delta T_{i\tau }=\delta T_{\tau i}=P\eta_{i\tau }-N\left( \rho+P \right)\left(\nabla_{i}\delta u + \delta u^{V}_{i}\right),& \nonumber \\[5pt] &\delta T_{ij}=P\eta_{ij}+a^{2}\tilde{\gamma}_{ij}\delta P+a^{2}\left( \nabla_{i}\nabla_{j}\pi^{S}+\nabla_{i}\pi^{V}_{j}+\nabla_{j}\pi^{V}_{i}+\pi^{T}_{ij}  \right),&
\label{eq:PerturbacionEnergiaMomentoCorrectDisipativas}
\end{eqnarray}
From the mixed perturbed energy-momentum tensor $T^{\mu}_{\ \nu}=\gamma^{\mu\lambda}T_{\lambda\nu}$, we can obtain the perturbed part
\begin{eqnarray}
\delta T^{\mu}_{\ \nu}=\gamma^{\mu\lambda}\left( \delta T_{\lambda\nu}-\eta_{\lambda\sigma}T^{\sigma}_{\ \nu} \right),
\end{eqnarray}
in which we have the components
\begin{eqnarray}
&\delta T^{\tau }_{\ \tau }=-\delta\rho , \quad \delta T^{i}_{\ j}=\delta^{i}_{j}\delta P, \quad \delta T^{\tau }_{\ i}=\frac{1}{N}\left( \rho+P \right)\delta u_{i} ,& \nonumber \\[5pt]
&\delta T^{i}_{\ \tau }=\frac{1}{a^{2}} \tilde{\gamma}^{im}\left[\left( \rho+P \right)\left(\eta_{m\tau }-N\delta u_{m}\right) \right],& \nonumber \\[5pt]
&\delta T^{\lambda}_{\ \lambda}=-\delta\rho+D\delta P.&
\end{eqnarray}

%%%%%%%%%%%%%%%%%%%%%%%%%%%%%%
\subsection{Scalar Modes}
%%%%%%%%%%%%%%%%%%%%%%%%%%%%%%%

In these modes the eight scalars $E$, $F$, $A$, $B$, $\delta\rho$, $\delta P$, $\pi^{S}$ and $\delta u$ are involved. The part of Eq. \eqref{eq:EcEinsteinTensor} proportional to $\tilde{\gamma}_{jk}$ gives
{\normalsize
\begin{eqnarray}
\frac{\kappa a^{2}}{D-1}\left[\delta \rho -\delta P - \tilde{\gamma}^{nm}\nabla_{n}\nabla_{m}\pi^{S}\right]&=&\left[ \left(D-1\right)\frac{\dot{a}^{2}}{N^{2}}+\frac{a\ddot{a}}{N^{2}}-\frac{a\dot{a}\dot{N}}{N^{3}} \right]E+\frac{1}{2}\frac{a\dot{a}}{N^{2}}\dot{E} \nonumber \\
& &+\frac{1}{2} \tilde{\gamma}^{nm}\nabla_{n}\nabla_{m}A+\left(\frac{a^{2}\dot{N}}{2N^{3}}-D\frac{a\dot{a}}{N^{2}}\right)\dot{A}-\frac{a^{2}}{2N^{2}}\ddot{A}  \nonumber\\  
& &-\frac{a\dot{a}}{2N^{2}}\tilde{\gamma}^{nm}\nabla_{n}\nabla_{m}\dot{B}+\frac{\dot{a}}{N}\tilde{\gamma}^{nm}\nabla_{n}\nabla_{m}F\nonumber\\&&+\left[\left(D-1\right)K+\frac{a\dot{a}\dot{N}}{N^{3}}-\frac{2}{D-1}\Lambda a^{2}  \right]A,
\label{eq:MScalar1}
\end{eqnarray}
}
additionally, the part of Eq. \eqref{eq:EcEinsteinTensor} of the form $\nabla_{j}\nabla_{k}S$ (where $S$ is any scalar) is
\begin{eqnarray}
& &\nabla_{j}\nabla_{k}\left\{- 2 \kappa a^{2} \pi^{S}+E+\left(D-2\right)A-\frac{a^{2}}{N^{2}}\ddot{B}+\left(\frac{a^{2}\dot{N}}{N^{3}}-D\frac{a\dot{a}}{N^{2}}\right)\dot{B}+2\frac{a}{N}\dot{F}\right. \nonumber \\
& &+\left. 2\left(D-1\right)\frac{\dot{a}}{N}F+\left[2\left(D-1\right)K+2\frac{a\dot{a}\dot{N}}{N^{3}}-\frac{4}{D-1}\Lambda a^{2}\right]B \right\}=0,
\label{eq:MScalar2}
\end{eqnarray}
for the part of Eq. \eqref{eq:EcEinsteinVector} of the form $\nabla_{j}S$ (where $S$ is any scalar) gives
\begin{eqnarray}
-\frac{2 \kappa N}{D-1}\left(\rho+P\right)\nabla_{j}\delta u=-\frac{\dot{a}}{a}\nabla_{j}E+\nabla_{j}\dot{A}+\left(2\frac{N}{a}K+\frac{2}{D-1}\frac{\dot{a}\dot{N}}{N^{2}}-\frac{4}{\left(D-1\right)^{2}}\Lambda aN \right)\nabla_{j}F, \nonumber \\
\label{eq:MScalar3}
\end{eqnarray}
for Eq. \eqref{eq:EcEinsteinScalar} we have
\begin{eqnarray}
&&\frac{\kappa N^{2}}{D-1} \left[\left(D-2\right) \delta\rho+D\delta P+\tilde{\gamma}^{ml}\nabla_{m}\nabla_{l} \pi^{S} \right]=\nonumber \\
&&-\frac{N^{2}}{2a^{2}}\tilde{\gamma}^{ml}\nabla_{m}\nabla_{l}E-\frac{D\dot{a}}{2a}\dot{E}-\frac{N}{a}\tilde{\gamma}^{ml}\nabla_{m}\nabla_{l}\dot{F} \nonumber \\
&&-\frac{\dot{a}}{a^{2}}N\tilde{\gamma}^{ml}\nabla_{m}\nabla_{l} F+\frac{D}{2}\ddot{A}-D\left(\frac{\ddot{a}}{a}-\frac{\dot{a}\dot{N}}{aN}+\frac{2}{D\left(D-1\right)}\Lambda N^{2}\right)E \nonumber \\
&&+\frac{1}{2}\tilde{\gamma}^{ml}\nabla_{m}\nabla_{l}\ddot{B}+\left(\frac{\dot{a}}{a}-\frac{\dot{N}}{2N}\right)\left(D\dot{A}+ \tilde{\gamma}^{ml}\nabla_{m}\nabla_{l} \dot{B}\right).
\label{eq:MScalar4}
\end{eqnarray}
Finally, the momentum conservation condition \eqref{eq:ConservacionMomento} which is a derivative $\nabla_{j}$ is
\begin{eqnarray}
\nabla_{j}\left[ \delta P+\tilde{\gamma}^{nm}\nabla_{n}\nabla_{m} \pi^{S}+\frac{1}{N}\partial_{\tau }\left[\left(\rho+P\right)\delta u\right]+D\frac{\dot{a}}{aN}\left(\rho+P\right)\delta u+\frac{1}{2}\left(\rho+P\right)E \right]=0,\nonumber \\
\label{eq:MScalar5}
\end{eqnarray}
and the energy conservation condition \eqref{eq:ConservacionEnergia} is
\begin{eqnarray}
& &\delta \dot{\rho}+\frac{D\dot{a}}{a}\left( \delta \rho + \delta P \right)+ \tilde{\gamma}^{ml}\nabla_{m}\nabla_{l}\left[\frac{N}{a}\left(\rho+P\right)\left(-F+a^{-1}\delta u\right) +\frac{\dot{a}}{a}\pi^{S} \right] \nonumber \\
& &+\frac{1}{2} \left(\rho+P\right) \partial_{\tau }\left[ DA+\tilde{\gamma}^{ml}\nabla_{m}\nabla_{l}B \right]=0.
\label{eq:MScalar6}
\end{eqnarray}
Notice that Eqs. \eqref{eq:MScalar5}, \eqref{eq:MScalar6}, $\delta \rho$, $\delta P$, and $\pi^{S}$ are elements of the perturbation to the total energy-momentum tensor, but the same equations apply to each constituent of the universe that does not exchange energy and momentum with other constituents.

%%%%%%%%%%%%%%%%%%%%%%%%%%%%%%%
\subsection{Vector Modes}
%%%%%%%%%%%%%%%%%%%%%%%%%%%%%%%

These modes that involve the divergenceless vectors are $G_{i}$, $C_{i}$, $\delta u^{V}_{i}$, and $\pi^{V}_{i}$. The part of Eq. \eqref{eq:EcEinsteinTensor} for the form $\nabla_{k}V_{j}$ (where $V_{j}$ is a divergenceless vector) give us
\begin{eqnarray}
& &\nabla_{k}\left\{2 \kappa a^{2} \pi^{V}_{j}+\left(D\frac{a\dot{a}}{N^{2}}-\frac{a^{2}\dot{N}}{N^{3}} \right)\dot{C}_{j}+\frac{a^{2}}{N^{2}}\ddot{C}_{j}-\left( D-1 \right)\frac{\dot{a}}{N}G_{j}\right. \nonumber \\
& &-\left.\frac{a}{N}\dot{\gamma}_{j}+\left[-2\left(D-1\right)K+\frac{4}{D-1}\Lambda a^{2}\right]C_{j} \right\}=0,
\label{eq:MVector1}
\end{eqnarray}
while the part of Eq. \eqref{eq:EcEinsteinVector} of the form $V_{j}$ (where $V_{j}$ is any divergenceless vector) reads
\begin{eqnarray}
-\kappa \left( \rho+P \right)N \delta u^{V}_{j}=&&\frac{1}{2}\frac{N}{a}\tilde{\gamma}^{ml}\nabla_{m}\nabla_{l}G_{j}-\frac{1}{2}\tilde{\gamma}^{ml}\nabla_{m}\nabla_{l}\dot{C}_{j}\nonumber \\
&&+\left[\left(D-1\right)K\frac{N}{a}+\frac{\dot{a}\dot{N}}{N^{2}}-\frac{2}{D-1}\Lambda aN\right]G_{j}.
\label{eq:MVector2}
\end{eqnarray}
The equation associated to the conservation of momentum \eqref{eq:ConservacionMomento} that takes the form of a divergenceless vector as
\begin{eqnarray}
\tilde{\gamma}^{lm}\nabla_{m}\nabla_{l} \pi^{V}_{j}+\frac{1}{N}\partial_{\tau }\left[ \left(\rho+P  \right)\delta u^{V}_{j} \right]+D\frac{\dot{a}}{aN}\left(\rho+P  \right)\delta u^{V}_{j}=0.
\label{eq:MVector3}
\end{eqnarray}
In particular, for a perfect fluid $\pi^{V}_{i}=0$, and Eq. \eqref{eq:MVector3} tells us that $\left(\rho+P  \right)\delta u^{V}_{j}$ decays as $a^{-D}$. In this case with $K=0$ and $\Lambda=0$, the Eq. \eqref{eq:MVector1} imply that the quantity $G_{i}-a\dot{C}_{i}/N$ decays as $a^{-D+1}$. Notice that they decay of vector modes do not play an important role in cosmology and therefore do not emerge from inflationary models.

%%%%%%%%%%%%%%%%%%%%%%%%%%%%%%
\subsection{Tensor Modes}
%%%%%%%%%%%%%%%%%%%%%%%%%%%%%%

These modes involve the 2-symmetric traceless and divergenceless tensors $D_{ij}$ and $\pi^{T}_{ij}$. The part of Eq. \eqref{eq:EcEinsteinTensor} of the form of a traceless and divergenceless tensor is the wave equation for gravitational radiation, given by the expression
\begin{eqnarray}
2\kappa a^{2} \pi^{T}_{ij}=&&\tilde{\gamma}^{ml}\nabla_{m}\nabla_{l}D_{ij}-\frac{a^{2}}{N^{2}}\ddot{D}_{ij}-\left(D\frac{a\dot{a}}{N^{2}}-\frac{a^{2}\dot{N}}{N^{3}}\right)\dot{D}_{ij}\nonumber \\
&&+2\left[\left( D-1 \right) K +\frac{a\dot{a}\dot{N}}{N^{3}}-\frac{2}{D-1}\Lambda a^{2}\right]D_{ij}.
\label{eq:ModoTensor}
\end{eqnarray}
Finally we have the condition of conservation of the current $\nabla_{\mu}(nu^{\mu})=0$. This condition tells us that the unperturbed density number complies $n \propto a^{-D}$, therefore from the perturbed part we have
\begin{eqnarray}
\frac{n}{2a^{2}N}\tilde{\gamma}^{lm}\dot{h}_{ml}+\frac{n}{N}\partial_{\tau }\left(\frac{\delta n}{n}\right)+\nabla_{l}\left(n\delta u^{l}\right)=0.
\label{eq:Corriente2}
\end{eqnarray}
From the previous equation when considering only the scalar part of the perturbation of the metric, we have
\begin{eqnarray}
\frac{1}{N}\partial_{\tau }\left(\frac{\delta n}{n}\right)+\frac{1}{2N}\left(D\dot{A}+\tilde{\gamma}^{lm}\nabla_{m}\nabla_{l}\dot{B}\right)+\frac{1}{a^{2}}\tilde{\gamma}^{lm}\nabla_{m}\nabla_{l}\left(\delta u-aF\right)=0. 
\label{eq:MScalar7}
\end{eqnarray}
Thus, equations \eqref{eq:MScalar5}, \eqref{eq:MScalar6} and \eqref{eq:MScalar7} form a complete set for scalar modes.

%%%%%%%%%%%%%%%%%%%%%%%%%%%%%%%%%%%%%%%%%%%%%%
\section{Gauges in Perturbations} \label{sec:Gauges}
%%%%%%%%%%%%%%%%%%%%%%%%%%%%%%%%%%%%%%%%%%%%%%

The solutions of the equations derived so far are non-physical scalar and vector modes, they correspond to a coordinate transformation of the metric and the energy-momentum tensor, to avoid this problem it is possible to adopt a suitable conditions for the perturbed metric and energy-momentum tensor. 

Let start considering a space-time coordinate transformation as \cite{weinbergcosmology} 
\begin{eqnarray}
x^{\mu}\rightarrow x'^{\mu}=x^{\mu}+\epsilon^{\mu}\left(x\right),
\label{eq:TransformacionCoordenada}
\end{eqnarray}
with $\epsilon^{\mu}\left(x\right)\ll1$. We work with the gauge transformations, which affect only the field perturbations. For this purpose, after making the coordinate transformation (\ref{eq:TransformacionCoordenada}), we relabel coordinates by dropping the prime on the coordinate argument, and we attribute the whole change in $\bar{\gamma}_{\mu\nu}(x)$ to a change in the perturbation $\eta_{\mu\nu}(x)$. The field equations should thus be invariant under the gauge transformation $\eta_{\mu\nu}\left(x\right)\rightarrow \eta_{\mu\nu}\left(x\right)+\Delta \eta_{\mu\nu}\left(x\right)$,
where
\begin{eqnarray}
\Delta \eta_{\mu\nu}\left(x\right)\equiv \bar{\gamma}'_{\mu\nu}(x)-\bar{\gamma}_{\mu\nu}(x).
\label{eq:Deltah}
\end{eqnarray} 
To first order in $\epsilon(x)$ and $\eta_{\mu\nu}(x)$, Eq. \eqref{eq:Deltah} in terms of the unperturbed metric is
\begin{eqnarray}
\Delta \eta_{\mu\nu}\left(x\right)&=&\bar{\gamma}'_{\mu\nu}(x')-\frac{\partial \bar{\gamma}_{\mu\nu}(x)}{\partial x^{\lambda}}\epsilon^{\lambda}\left(x \right)-\bar{\gamma}_{\mu\nu}(x)\nonumber \\
&=&-\gamma_ {\lambda\mu}(x)\frac{\partial \epsilon^{\lambda}\left(x\right)}{\partial x^{\nu}}-\gamma_ {\lambda\nu}\left(x\right)\frac{\partial \epsilon^{\lambda}\left(x\right)}{\partial x^{\mu}}-\frac{\partial \gamma_ {\mu\nu}(x)}{\partial x^{\lambda}} \epsilon^{\lambda}\left(x\right),
\label{eq:Deltah1}
\end{eqnarray}
this applied to the FLRW metric
\begin{eqnarray}
& &\Delta \eta_{\tau \tau }=2\frac{\dot{N}}{N}\epsilon_{\tau }-2\frac{\partial \epsilon_{\tau }}{\partial t}, \nonumber \\[5pt]
& &\Delta \eta_{\tau i}=-\frac{\partial \epsilon_{i}}{\partial t}-\nabla_{i}\epsilon_{\tau }+2\frac{\dot{a}}{a}\epsilon_{i},\nonumber \\[5pt]
& &\Delta \eta_{ij}=-\nabla_{j}\epsilon_{i}-\nabla_{i}\epsilon_{j}+2\frac{a\dot{a}}{N^{2}}\tilde{\gamma}_{ij}\epsilon_{\tau },
\label{eq:Transformacionesh}
\end{eqnarray}
with all quantities evaluated at the same space-time coordinate point. The field equations will be invariant only if the same gauge transformation is applied to all tensors, and in particular to the energy-momentum tensor. A more generalized formula analogous to Eq. \eqref{eq:Deltah1} is
\begin{eqnarray}
\Delta \delta Z_{\mu\nu}\left(x\right)&=&-Z_{\lambda\mu}(x)\frac{\partial \epsilon^{\lambda}\left(x\right)}{\partial x^{\nu}}-Z_{\lambda\nu}\left(x\right)\frac{\partial \epsilon^{\lambda}\left(x\right)}{\partial x^{\mu}}-\frac{\partial Z_{\mu\nu}(x)}{\partial x^{\lambda}} \epsilon^{\lambda}\left(x\right),
\label{eq:DeltaZ1}
\end{eqnarray}
or explicitly for FLRW

\begin{eqnarray}
& &\Delta \delta Z_{\tau \tau }=-2\frac{\dot{N}}{N}\zeta\epsilon_{\tau }+2\zeta\frac{\partial \epsilon_{\tau }}{\partial t}+\dot{\zeta}\epsilon_{\tau }, \nonumber \\[5pt]
& &\Delta \delta Z_{\tau i}=-\xi\frac{\partial \epsilon_{i}}{\partial t}+\zeta\nabla_{i}\epsilon_{\tau }+2\xi\frac{\dot{a}}{a}\epsilon_{i},\nonumber \\[5pt]
& &\Delta \delta Z_{ij}=-\xi\left(\nabla_{j}\epsilon_{i}+\nabla_{i}\epsilon_{j}\right)+\frac{1}{N^{2}}\frac{\partial}{\partial t}\left(a^{2}\xi\right)\tilde{\gamma}_{ij}\epsilon_{\tau } .
\label{eq:TransformacionesZ}
\end{eqnarray}
This applies to a tensor of second order, with components
\begin{eqnarray}
Z_{\tau \tau }=\zeta(t)N^{2}; \quad\quad Z_{\tau i}=Z_{i\tau }=0; \quad\quad Z_{ij}=\xi(t)a^{2}\tilde{\gamma}_{ij}.
\end{eqnarray}
Note that the transformations \eqref{eq:Deltah1} and \eqref{eq:DeltaZ1} are for any arbitrary metric and second-order tensor. Note also, that we use $\delta$ for a perturbation, while $\Delta$ denotes the change in a perturbation associated with a gauge transformation. In particular, for the energy-momentum tensor of a perfect fluid $\zeta=\rho$ and $\xi=P$.

In order to write these gauge transformations in terms of the scalar, vector, and tensor components, it is necessary to decompose the spatial part of $\epsilon^{\mu}$ into the gradient of a spatial scalar, plus a divergenceless vector, in the form
\begin{eqnarray}
\epsilon_{i}=\nabla_{i}\epsilon^{S}+\epsilon^{V}_{i}, \quad\quad \nabla_{i}\epsilon_{i}^{V}=0.
\end{eqnarray}
Thus the transformations \eqref{eq:Transformacionesh} and \eqref{eq:TransformacionesZ} give us the gauge transformations of the metric perturbation components defined by Eqs. \eqref{eq:Descomposicion}

\begin{eqnarray}
&\Delta E=\frac{2}{N^{2}}\left(\dot{\epsilon}_{\tau }-\frac{\dot{N}}{N}\epsilon_{\tau }\right), \quad \Delta F=\frac{1}{aN}\left(-\epsilon_{\tau }-\dot{\epsilon}^{S}+2\frac{\dot{a}}{a}\epsilon^{S}\right), \quad \Delta G_{i}=\frac{1}{aN}\left(-\dot{\epsilon}_{i}^{V}+2\frac{\dot{a}}{a}\epsilon_{i}^{V}\right),& \nonumber \\[5pt]
&\Delta A=2\frac{\dot{a}}{aN^{2}}\epsilon_{\tau }, \quad \Delta B=-\frac{2}{a^{2}}\epsilon^{S}, \quad \Delta C_{i}=-\frac{1}{a^{2}}\epsilon_{i}^{V}, \quad \Delta D_{ij}=0,&
\end{eqnarray}
and from \eqref{eq:PerturbacionEnergiaMomentoCorrectDisipativas} for the pressure, energy density, and velocity potential, reads
\begin{eqnarray}
\Delta\delta\rho=\frac{\dot{\rho}\epsilon_{\tau }}{N^{2}}, \quad \Delta\delta P= \frac{\dot{P}\epsilon_{\tau }}{N^{2}}, \quad \Delta\delta u=-\frac{\epsilon_{\tau }}{N}.
\label{eq:TGaugesRhoP}
\end{eqnarray}
The other ingredients of the energy-momentum tensor are gauge invariant
\begin{eqnarray}
\Delta \pi^{S}=\Delta \pi^{i}_{V}=\Delta \pi^{ij}_{T}=\Delta\delta u^{V}_{i}=0.
\label{eq:GuageInvariant}
\end{eqnarray}
Note in particular that the conditions $\pi^{S}=\pi^{i}_{V}=\pi^{ij}_{T}=0$ for a perfect fluid and the condition $\delta u^{V}_{i}=0$ for the potential flow are gauge invariant. In general, any space-time-scalar $\vartheta(x)$ for which $\vartheta '(x')=\vartheta(x)$ under arbitrary coordinate transformations would undergo the change $\Delta \delta \vartheta (x)\equiv \vartheta '(x)-\vartheta(x)= \vartheta '(x)-\vartheta '(x')$, which to first order in perturbations is
\begin{eqnarray}
\Delta \delta \vartheta (x) = \vartheta (x)-\vartheta (x') = -\frac{\partial \vartheta (t)}{\partial x^{\mu}}\epsilon^{\mu}(x)=\frac{\dot{\vartheta} (t)}{N^{2}}\epsilon_{\tau }.
\label{eq:TGaugesScalar}
\end{eqnarray}
This applies for instance to the number density $n$ or a scalar field $\varphi$. For a perfect fluid both $\rho$ and $P$ are defined as scalars, and the gauge transformations in Eq. \eqref{eq:TGaugesRhoP} of $\delta\rho$ and $\delta P$ are other special cases of Eq. \eqref{eq:TGaugesScalar}. Likewise, for a perfect fluid the gauge transformation in Eq. \eqref{eq:TGaugesRhoP} of $\delta u$ can be derived from the vector transformation law of $u_{\mu}$. Because the gauge transformation properties of $\delta\rho$, $\delta P$, $\delta u$, etc. do not depend on the conservation laws, Eqs. \eqref{eq:TGaugesRhoP} and \eqref{eq:GuageInvariant} apply to each individual constituent of the universe in any case in which the energy-momentum tensor is a sum of terms for different constituents of the universe, even if these individual terms are not separately conserved. 

It is possible to neglect the gauge degrees of freedom either by working only with gauge-invariant quantities, or by choosing a gauge. The tensor quantities $\pi_{ij}$ and $D_{ij}$ appearing in Eq. \eqref{eq:ModoTensor} are already gauge invariant, and non gauge-fixing is necessary or possible. For the vector quantities $\pi^{V}_{i}$, $\delta u^{V}_{i}$, $C_{i}$ and $G_{i}$, we can write Eqs. \eqref{eq:MVector1}-\eqref{eq:MVector3} in terms of the gauge-invariant quantities $\pi^{V}_{i}$, $\delta u^{V}_{i}$ and $G_{i}-(a/N)\dot{C}_{i}$, or we can fix a gauge for these quantities by choosing $\epsilon^{V}_{i}$, so that either $C_{i}$ or $G_{i}$ vanishes. For the scalar perturbations it is somewhat more convenient to fix a gauge.

%%%%%%%%%%%%%%%%%%%%%%%%%%%%%%%%%%%%%%%%%%%%%%%%
\subsection{Newtonian Conformal Gauge}
%%%%%%%%%%%%%%%%%%%%%%%%%%%%%%%%%%%%%%%%%%%%%%%%

In this gauge only scalar perturbations are considered \cite{weinbergcosmology,dodelsoncosmology,Ma_1995}. The line element of such gauge is
\begin{eqnarray}
ds^{2}= -N^{2}\left( 1+2\Phi \right)dt^{2}+a^{2}(1-2\Psi)\tilde{\gamma}_{ij}dx^{i}dx^{j}, 
\end{eqnarray}
here we choose $\epsilon^{S}$ so that $B=0$, and then choose $\epsilon_{\tau }$ so that $F=0$. Both choices are unique, so that after choosing Newtonian gauge, there is no remaining freedom to make gauge transformations. Thus in this gauge, we have $A=-2\Psi$, $E=2\Phi$, $B=0$ and $F=0$.

The gravitational field equations \eqref{eq:MScalar1}-\eqref{eq:MScalar4}, takes the form
{\normalsize
\begin{eqnarray}
\frac{\kappa a^{2}}{D-1}\left[\delta \rho -\delta P - \tilde{\gamma}^{nm}\nabla_{n}\nabla_{m}\pi^{S}\right]&=&\left[ \left(D-1\right)\frac{\dot{a}^{2}}{N^{2}}+\frac{a\ddot{a}}{N^{2}}-\frac{a\dot{a}\dot{N}}{N^{3}} \right]2\Phi+\frac{a\dot{a}}{N^{2}}\dot{\Phi} \nonumber \\
& &-\tilde{\gamma}^{nm}\nabla_{n}\nabla_{m}\Psi-\left(\frac{a^{2}\dot{N}}{2N^{3}}-D\frac{a\dot{a}}{N^{2}}\right)2\dot{\Psi}+\frac{a^{2}}{N^{2}}\ddot{\Psi}  \nonumber\\  
& &-\left[\left(D-1\right)K+\frac{a\dot{a}\dot{N}}{N^{3}}-\frac{2}{D-1}\Lambda a^{2}\right]2\Psi,
\label{eq:Newton1}
\end{eqnarray}
}

\begin{eqnarray}
\nabla_{j}\nabla_{k}\left\{-\kappa a^{2} \pi^{S}+\Phi-\left(D-2\right)\Psi\right\}=0,
\label{eq:Newton2}
\end{eqnarray}

\begin{eqnarray}
-\frac{ \kappa N}{D-1}\left(\rho+P\right)\nabla_{j}\delta u=-\frac{\dot{a}}{a}\nabla_{j}\Phi-\nabla_{j}\dot{\Psi},
\label{eq:Newton3}
\end{eqnarray}

\begin{eqnarray}
&&\frac{\kappa N^{2}}{D-1} \left[\left(D-2\right) \delta\rho+D\delta P+\tilde{\gamma}^{ml}\nabla_{m}\nabla_{l} \pi^{S} \right]=-\frac{N^{2}}{a^{2}}\tilde{\gamma}^{ml}\nabla_{m}\nabla_{l}\Phi-\frac{D\dot{a}}{a}\dot{\Phi}\nonumber \\
&&-D\left(\frac{\ddot{a}}{a}-\frac{\dot{a}\dot{N}}{aN} +\frac{2}{D\left(D-1\right)}\Lambda N^{2} \right)2\Phi-D\ddot{\Psi}-D\left(2\frac{\dot{a}}{a}-\frac{\dot{N}}{N}\right)\dot{\Psi}.
\label{eq:Newton4}
\end{eqnarray}
The conservation conditions \eqref{eq:MScalar5}, \eqref{eq:MScalar6} and \eqref{eq:MScalar7} becomes
\begin{eqnarray}
\delta P+\tilde{\gamma}^{nm}\nabla_{n}\nabla_{m} \pi^{S}+\frac{1}{N}\partial_{\tau }\left[\left(\rho+P\right)\delta u\right]+D\frac{\dot{a}}{aN}\left(\rho+P\right)\delta u+\left(\rho+P\right)\Phi =0,
\label{eq:Newton5}
\end{eqnarray}

\begin{eqnarray}
& &\delta \dot{\rho}+\frac{D\dot{a}}{a}\left( \delta \rho + \delta P \right)+ \tilde{\gamma}^{ml}\nabla_{m}\nabla_{l}\left[\frac{N}{a^{2}}\left(\rho+P\right)\delta u +\frac{\dot{a}}{a}\pi^{S} \right] -D\left(\rho+P\right)\dot{\Psi} =0,
\label{eq:Newton6}
\end{eqnarray}

\begin{eqnarray}
\frac{1}{N}\partial_{\tau }\left(\frac{\delta n}{n}\right)+\frac{1}{a^{2}}\tilde{\gamma}^{lm}\nabla_{m}\nabla_{l}\delta u-\frac{D}{N}\dot{\Psi}=0.
\label{eq:Newton7}
\end{eqnarray}
By performing algebra in the previous equations we arrive to the following constraint

\begin{eqnarray}
a^{3} \delta\rho-DH\frac{a^{3}}{N}\left(\rho+P\right)\delta u + \left(D-1\right)\frac{a}{\kappa}\tilde{\gamma}^{lm}\nabla_{m}\nabla_{l}\Psi + \frac{D}{\kappa}\left[\left(D-1\right)Ka+\frac{a^{2}\dot{a}\dot{N}}{N^{3}}  \right]\Psi=0.\nonumber \\
\label{eq:NewtonConstriccion}
\end{eqnarray}

%%%%%%%%%%%%%%%%%%%%%%%%%%%%%%%%%%%%%%%%%%%%
\subsection{Synchronous gauge}
%%%%%%%%%%%%%%%%%%%%%%%%%%%%%%%%%%%%%%%%%%%%

In the synchronous gauge we have $A\neq 0$, $B\neq 0$, $E=0$ and $F=0$. The line element of such a gauge is
\begin{eqnarray}
ds^{2}=-N^{2}dt^{2}+a^{2}\left[(1+A)\tilde{\gamma}_{ij}+\nabla_{i}\nabla_{j}B\right]dx^{i}dx^{j}.
\end{eqnarray}
Hence, the gravitational field equations \eqref{eq:MScalar1}-\eqref{eq:MScalar4}, for this gauge takes the form
{\normalsize
\begin{eqnarray}
&&\frac{\kappa a^{2}}{D-1}\left[\delta \rho -\delta P - \tilde{\gamma}^{nm}\nabla_{n}\nabla_{m}\pi^{S}\right]=\frac{1}{2} \tilde{\gamma}^{nm}\nabla_{n}\nabla_{m}A+\left(\frac{a^{2}\dot{N}}{2N^{3}}-D\frac{a\dot{a}}{N^{2}}\right)\dot{A}-\frac{a^{2}}{2N^{2}}\ddot{A}  \nonumber\\  
& &-\frac{a\dot{a}}{2N^{2}}\tilde{\gamma}^{nm}\nabla_{n}\nabla_{m}\dot{B}+\left[\left(D-1\right)K+\frac{a\dot{a}\dot{N}}{N^{3}} -\frac{2}{D-1}\Lambda a^{2} \right]A, 
\label{eq:Syncro1}
\end{eqnarray}
}

\begin{eqnarray}
&&- 2 \kappa a^{2} \pi^{S}+\left(D-2\right)A-\frac{a^{2}}{N^{2}}\ddot{B}+\left(\frac{a^{2}\dot{N}}{N^{3}}-D\frac{a\dot{a}}{N^{2}}\right)\dot{B}\nonumber \\&&+\left[2\left(D-1\right)K+2\frac{a\dot{a}\dot{N}}{N^{3}} -\frac{4}{D-1}\Lambda a^{2}  \right]B=0,
\label{eq:Syncro2}
\end{eqnarray}

\begin{eqnarray}
-\frac{2 \kappa N}{D-1}\left(\rho+P\right)\delta u=\dot{A},
\label{eq:Syncro3}
\end{eqnarray}

\begin{eqnarray}
&&\frac{\kappa N^{2}}{D-1} \left[\left(D-2\right) \delta\rho+D\delta P+\tilde{\gamma}^{ml}\nabla_{m}\nabla_{l} \pi^{S} \right]=\nonumber \\
&&\frac{D}{2}\ddot{A}+\frac{1}{2}\tilde{\gamma}^{ml}\nabla_{m}\nabla_{l}\ddot{B}+\left(\frac{\dot{a}}{a}-\frac{\dot{N}}{2N}\right)\left(D\dot{A}+ \tilde{\gamma}^{ml}\nabla_{m}\nabla_{l} \dot{B}\right).
\label{eq:Syncro4}
\end{eqnarray}
The conservation conditions given by Eqs.  \eqref{eq:MScalar5}, \eqref{eq:MScalar6} and \eqref{eq:MScalar7} becomes
\begin{eqnarray}
\delta P+\tilde{\gamma}^{nm}\nabla_{n}\nabla_{m} \pi^{S}+\frac{1}{N}\partial_{\tau }\left[\left(\rho+P\right)\delta u\right]+D\frac{\dot{a}}{aN}\left(\rho+P\right)\delta u =0,
\label{eq:Syncro5}
\end{eqnarray}

\begin{eqnarray}
& &\delta \dot{\rho}+\frac{D\dot{a}}{a}\left( \delta \rho + \delta P \right)+ \tilde{\gamma}^{ml}\nabla_{m}\nabla_{l}\left[\frac{N}{a^{2}}\left(\rho+P\right)\delta u +\frac{\dot{a}}{a}\pi^{S} \right] \nonumber \\
& &+\frac{1}{2} \left(\rho+P\right) \partial_{\tau }\left[ DA+\tilde{\gamma}^{ml}\nabla_{m}\nabla_{l}B \right]=0,
\label{eq:Syncro6}
\end{eqnarray}

\begin{eqnarray}
\frac{1}{N}\partial_{\tau }\left(\frac{\delta n}{n}\right)+\frac{1}{2N}\left(D\dot{A}+\tilde{\gamma}^{lm}\nabla_{m}\nabla_{l}\dot{B}\right)+\frac{1}{a^{2}}\tilde{\gamma}^{lm}\nabla_{m}\nabla_{l}\delta u=0. 
\label{eq:Syncro7}
\end{eqnarray}
Before finish, we remark that all equations shown in this part of the study recover the successes of GR in the limit where $D\to3$, having the standard $3+1$ space-time structure.

%%%%%%%%%%%%%%%%%%%%%%%%%%%%%%%%%%%%%%%%%%%%%%%
\section{Branes perturbations: A brane embedded in a $D + 1 + 1$ bulk} \label{sec:PertCosmoBrane} 
%%%%%%%%%%%%%%%%%%%%%%%%%%%%%%%%%%%%%%%%%%%%%%%%%

Our configuration is a bulk of dimension $M$, in this space-time there is a embedded purely spatial hypersurface of dimension $D$, whose coordinate system is $\left( ...x^{i},x^{j},x^{k}... \right)$, there is one extra spatial dimension, whose coordinate system is $a$, and one temporal dimension. So $M=D+1+1$ and the coordinate system for the bulk is $\left(t,a,x^{i},x^{j},x^{k}... \right)$ \cite{Cordero_2002,Garc_a_Aspeitia_2010}. Let us consider that in our bulk has a cosmological constant. The gravitational interaction inside the brane is not taken into account, so we have the following action
\begin{eqnarray}
S=S_{gravity}+S_{brane},
\end{eqnarray}
where
\begin{eqnarray}
S_{gravity}=\int dt\;da\;d^{D}x\frac{\sqrt{-g}}{\kappa_{B}}\left( -\Lambda+\frac{1}{2}R \right),
\end{eqnarray}

\begin{eqnarray}
S_{brane}=\int dt\;d^{D}x\sqrt{-\gamma}\left( \varepsilon\frac{1}{\kappa}\left\langle trK \right\rangle+\mc{L} \right).
\end{eqnarray}
In the bulk we have that, $g$ is the determinant of the metric, $R$ is the scalar of curvature of
Ricci and $\kappa_{B}\propto G_{B}$, $G_{B}$ is a gravitational constant for bulk. On the brane we have to,
$\gamma$ is the determinant of the metric, $trK$ is the
trace of the extrinsic curvature $K_{\mu\nu}$, $\mc{L}$ es the Lagrangian of the content of the fields on the brane,
$\varepsilon$ is $+1$ if the brane is {\it temporalloid} and $-1$ if it is {\it spatialloid}, for our case $\varepsilon=-1$, the bracket
$\left\langle trK \right\rangle=trK^{+}-trK^{-}$, it would be the same for all objects, the subscript $+$ or $-$ indicates whether it is
considering the metric above or below the brane respectively. Notice that GR in four dimensional space-time is recovered in all the equations when brane does not exist implying null extrinsic curvature and $D=2$ for this particular case. Thus we recover GR with $3+1$ space-time manifold, specifying that it would only be maximally symmetric in two spatial dimensions.

The line element in the bulk is,
\begin{eqnarray}
ds^{2}&=&g_{00}dt^{2}+g_{aa}da^{2}+g_{ij}dx^{i}dx^{j} \nonumber \\
&=&-\mc{A}_{\pm}dt^{2}_{\pm}+\mc{B}_{\pm}da^{2}+\mc{C}\tilde{g}_{ij}dx^{i}dx^{j}, 
\label{eq:MetricaBulk}
\end{eqnarray} 
where $\mc{A}_{\pm},\, \mc{B}_{\pm},$ and $\mc{C}$ are functions of the scale factor $a$, furthermore, $g_{ta}=g_{ti}=g_{ai}=0$. The metric for space-time complies with $g_{ML}=g_{LM}$, where $M, \,L=t,\,a,\,,\,i,\,j,\,k,\,...$.

The line element in the brane, is
\begin{eqnarray}
ds^{2}&=&\gamma_{\tau\tau}d\tau^{2}+\gamma_{ij}dx^{i}dx^{j}=-N^{2}d\tau^{2}+a^{2}\tilde{\gamma}_{ij}dx^{i}dx^{j}, 
\label{eq:MetricaBrana}
\end{eqnarray} 
where $N$ and $a$ are functions of the time $\tau$ (proper time of the brane). The spatial hypersurface is endowed with an arbitrary sectional curvature $K$ and of dimension $D$, which is the same metric as in section \ref{sec:PertCosmoStandar}. $\mc{A}_{\pm}$ and $\mc{B}_{\pm}$ are also functions of $K$. Notice that the temporal coordinate for the bulk is $t$ and for the brane is $\tau$. 

With this we indicate that the capital Latin indices are for the bulk, the Greek indices are for the brane and the small Latin indices are for the purely spatial hypersurface. 

Note that the line element on the brane must match that of the bulk, i.e.
\begin{eqnarray}
ds^{2}_{brane}=ds^{2}_{Bulk},
\end{eqnarray}
For the particular case of this paper, the spatial coordinates of the brane are comoving and the extra and temporal dimension of the bulk are parameterized by the temporal coordinate of the brane, therefore
\begin{eqnarray}
-N^{2}d \tau^{2}=-\mc{A}_{\pm}dt^{2}+\mc{B}_{\pm}da^{2}, \quad {\rm and} \quad a^{2}\tilde{\gamma}_{ij}dx^{i}dx^{j}=\mc{C}\tilde{g}_{ij}dx^{i}dx^{j}.
\end{eqnarray}
We modeled the linear perturbations in this way, because we focus on brane dynamics in the bulk and how this contributes internally to the brane, without considering internal brane dynamics beforehand.

%%%%%%%%%%%%%%%%%%%%%%%%%%%%%%%%%%%%%%%%%%%
\subsection{Junction Conditions}
%%%%%%%%%%%%%%%%%%%%%%%%%%%%%%%%%%%%%%%%%%%

The junction conditions for unperturbed spacetime have already been studied in the literature, we now want to impose such conditions for perturbed spacetime. Suppose that a continuous coordinate system $y^{A}$, distinct from $y_{A}^{\pm}$, can be introduced on both sides of the Brane.

Let us think that a congruence of geodesics of the bulk crosses orthogonally the brane. We have a scalar field $\mcb{l}$ along the geodesics, and we will adjust the parameterization such that $\mcb{l}=0$ when the geodesics cross the brane; $\mcb{l}$ will be considered positive or negative for the region above or below the brane respectively. The displacements along the geodesic away from the brane are described by $dy^{A}=n^{A}d\mcb{l}$, and that\cite{poisson2004toolkitl}

\begin{eqnarray}
n_{A}=\epsilon \partial_{A}\mcb{l},
\label{eq:normalwithl}
\end{eqnarray} 
with $\epsilon=n_{A}n^{A}$. Thus we have the bulk metric $g_{AB}$ of the system $y^{A}$, expressed as a distribution:

\begin{eqnarray}
g_{AB}=\Theta\left(\mcb{l}\right)g_{AB}^{+}+\Theta\left(-\mcb{l}\right)g_{AB}^{-},
\label{eq:MetricaDistribution}
\end{eqnarray} 
where $\Theta\left(l\right)$ is the Heaviside step function. We want to know if the metric of equation \eqref{eq:MetricaDistribution} makes a valid distributive solution for the Einstein field equations. Differentiating equation \eqref{eq:MetricaDistribution} we obtain

\begin{eqnarray}
g_{AB,C}=\Theta\left(\mcb{l}\right)g_{AB,C}^{+}+\Theta\left(-\mcb{l}\right)g_{AB,C}^{-}+\epsilon \delta\left(\mcb{l}\right)\left\langle g_{AB}\right\rangle n_{C},
\label{eq:DerivativeMetricaDistribution}
\end{eqnarray} 
for this we use the equation \eqref{eq:normalwithl}, and $\delta\left(\mcb{l}\right)$ is the Dirac delta function. The last term is singular and causes problems when we calculate the Christoffel symbols, and consequently all the geometric quantities derived from them. To eliminate this term we impose continuity of the metric across the brane: $\left\langle g_{AB}\right\rangle = 0$. This statement holds in the $y_{A}$ coordinate system only. However, we can easily convert this into a coordinate invariant statement: $0=\left\langle g_{AB}\right\rangle \xi^{A}_{\alpha}\xi^{B}_{\beta} =\left\langle g_{AB} \xi^{A}_{\alpha}\xi^{B}_{\beta}\right\rangle$; this is due to the condition

\begin{eqnarray}
\left\langle  \xi^{A}_{\alpha}\right\rangle =0,
\label{eq:ConditionMetricInduced}
\end{eqnarray} 
where $\xi^{A}_{\alpha}=\frac{\partial y^{A}}{\partial x^{\alpha}}$ and $x^{\alpha}$ is the brane coordinate system. So we have to

\begin{eqnarray}
\left\langle  \gamma_{\alpha\beta}\right\rangle =0,
\label{eq:FirstCondition}
\end{eqnarray} 
in order to have the claim that the induced metric must be the same on both sides of the brane. This is necessary for the brane to have a well-defined geometry. The equation \eqref{eq:FirstCondition} will be our first join condition and is expressed regardless of the coordinates $y^{A}$ or $y^{A}_{\pm}$.

The same treatment can be applied to a perturbed space-time and we would obtain $\left\langle \bar{\gamma}_{\alpha\beta}\right\rangle =0$, and as a consequence

\begin{eqnarray}
\left\langle  \eta_{\alpha\beta}\right\rangle =0,
\label{eq:FirstConditionPerturbed}
\end{eqnarray} 
this is the first junction condition for the perturbation of the metric.

To find the second junction condition we must calculate the Riemann tensor with distribution value. Using the results obtained so far, we have that the Christoffel symbols are

\begin{eqnarray}
\Gamma^{A}_{BC}=\Theta\left(\mcb{l}\right)\Gamma^{+A}_{BC}+\Theta\left(-\mcb{l}\right)\Gamma^{-A}_{BC},
\label{eq:ChristoffelDistribution}
\end{eqnarray} 
differentiating we have

\begin{eqnarray}
\Gamma^{A}_{BC,D}=\Theta\left(\mcb{l}\right)\Gamma^{+A}_{BC,D}+\Theta\left(-\mcb{l}\right)\Gamma^{-A}_{BC,D}+\epsilon \delta\left(\mcb{l}\right)\left\langle \Gamma^{A}_{BC}\right\rangle n_{D},
\label{eq:DerivativeChristoffelDistribution}
\end{eqnarray} 
and from this follows the Riemann tensor

\begin{eqnarray}
R^{A}_{BCD}=\Theta\left(\mcb{l}\right)R^{+A}_{BCD}+\Theta\left(-\mcb{l}\right)R^{-A}_{BCD}+\delta\left(\mcb{l}\right)\left\langle A^{A}_{BCD}\right\rangle,
\label{eq:ChristoffelDistribution}
\end{eqnarray}
where

\begin{eqnarray}
A^{A}_{BCD}=\epsilon\left(\left\langle \Gamma^{A}_{BD}\right\rangle n_{C}-\left\langle \Gamma^{A}_{BC}\right\rangle n_{D}\right).
\label{eq:ChristoffelDistribution}
\end{eqnarray}
With this we form the part of the $\delta-$function of the Einstein tensor, and after using the Einstein field equations we obtain an expression for the energy-momentum tensor

\begin{eqnarray}
\mcb{T}_{AB}=\Theta\left(\mcb{l}\right)\mcb{T}_{AB}^{+}+\Theta\left(-\mcb{l}\right)\mcb{T}_{AB}^{-}+\delta\left(\mcb{l}\right)T_{AB},
\label{eq:Stress-EnergyDistribution}
\end{eqnarray} 
where, $\kappa T_{AB} \equiv A_{AB}-\frac{1}{2}Ag_{AB}$. The term of the $\delta-$function is associated with the presence of a fine distribution of matter in the brane. Supporting us with

\begin{eqnarray}
\left\langle n_{A;B}\right\rangle = -\left\langle \Gamma^{C}_{AB}  \right\rangle n_{C},
\label{eq:DerivNormal=Gamma}
\end{eqnarray} 
allows us to write

\begin{eqnarray}
\left\langle K_{\alpha\beta}\right\rangle=\left\langle n_{A;B}\right\rangle \xi^{A}_{\alpha}\xi^{B}_{\beta},
\label{eq:KurvExt=Normal}
\end{eqnarray} 
and we get

\begin{eqnarray}
\left\langle K \right\rangle\gamma_{\mu\nu}-\left\langle K_{\mu\nu}\right\rangle=\epsilon\kappa T_{\mu\nu},
\label{eq:EinsteinHilbertCurvaturaExt}
\end{eqnarray}
where $T_{\mu\nu}=\xi^{M}_{\mu}\xi^{N}_{\nu}T_{MN}$\cite{dodelsoncosmology}. We conclude that a smooth transition requires $\left\langle K_{\mu\nu}\right\rangle = 0$, this eliminates the term of the $\delta$-function of the Einstein tensor and also implies that the Riemann tensor of the Bulk is non-singular on the brane.

This is the second condition, and it is expressed independently of the coordinates $y^{\alpha}$ or $y^{\alpha}_{\pm}$. If this condition is not met, then space-time is singular on the brane and in turn is provided with matter. In this case we can use the equation \eqref{eq:EinsteinHilbertCurvaturaExt} to study the dynamics of the brane.

As we mentioned before, this same treatment can be applied to a perturbed spacetime, thus having $\left\langle \bar{K}_{\mu\nu}\right\rangle = 0$ and therefore we have the version of the second condition for the perturbed part

\begin{eqnarray}
\left\langle \delta K_{\mu\nu}\right\rangle = 0,
\label{eq:SecondConditionPerturbed}
\end{eqnarray}
similarly, if this condition is not met, then the perturbed part of the brane is made up of matter, and the equation that describes the dynamics for perturbations is: 

\begin{eqnarray}
\left\langle \delta K \right\rangle\gamma_{\mu\nu}+\left\langle K \right\rangle\eta_{\mu\nu}-\left\langle \delta K_{\mu\nu}\right\rangle=\epsilon\kappa \delta T_{\mu\nu}.
\label{eq:EinsteinHilbertCurvaturaExt}
\end{eqnarray}

%%%%%%%%%%%%%%%%%%%%%%%%%%%%%%%%%%%%%%%%%%%
\subsection{Dynamic Equations}
%%%%%%%%%%%%%%%%%%%%%%%%%%%%%%%%%%%%%%%%%%%

Therefore, from the previous subsection and considering $\epsilon=1$ , the equation of motion of the branes is
\begin{eqnarray}
\left\langle K \right\rangle\gamma_{\mu\nu}-\left\langle K_{\mu\nu}\right\rangle=\kappa T_{\mu\nu},
\label{eq:EinsteinHilbertCurvaturaExt}
\end{eqnarray}
equation \eqref{eq:EinsteinHilbertCurvaturaExt} can be rewritten in terms of the source tensor $\left\langle K_{\mu\nu} \right\rangle=\kappa S_{\mu\nu}$, \cite{weinbergcosmology}
where
\begin{eqnarray}
S_{\mu\nu}=-T_{\mu\nu}+D^{-1}T^{\lambda}_{\ \lambda}\gamma_{\mu\nu}.
\end{eqnarray}
In addition, the extrinsic curvature is defined as
\begin{eqnarray}
K_{\mu\nu}=-\xi^{A}_{\mu}\xi^{B}_{\nu}\nabla_{B}n_{A}=-\xi^{A}_{\mu}\partial_{\nu}n_{A}+\xi^{A}_{\mu}\xi^{B}_{\nu}n_{C}\Gamma^{C}_{AB},
\label{eq:CurvaturaExtrinsica}
\end{eqnarray}  
The subscript $\pm$ will be omitted, and it will be shown explicitly when necessary, however it must be
understand that every quantity that depends on $\mc{A}, \,\mc{B}$, depends at the same time on the subscript $\pm$.

Taking into account that $\xi^{0}_{\tau}=\dot{t}$ and $\xi^{a}_{\tau}=\dot{a}$, we find that the extrinsic curvatures are
\begin{eqnarray}
K^{\pm}_{\tau\tau}=&&\mp\frac{N}{2\sqrt{\mc{AB}}\dot{t}}\left(\dot{t}^{2}\partial_{a}\mc{A}+\dot{a}^{2}\partial_{a}\mc{B}+2\mc{B}\left(\ddot{a}-\frac{\dot{a}\dot{N}}{N^{2}}\right)\right.\nonumber \\&&+\left.\frac{\dot{a}^{2}\dot{t}^{2}}{N^{2}}\left(1-N\right)\left(\mc{B}\partial_{a}\mc{A}+\mc{A}\partial_{a}\mc{B}\right)\right),
\label{eq:CurvaturaExtrinsicatt}
\end{eqnarray}

\begin{eqnarray}
K^{\pm}_{ij}=\pm\frac{\dot{t}}{2N}\sqrt{\frac{\mc{A}}{\mc{B}}}\tilde{\gamma}_{ij}\partial_{a}\mc{C},
\label{eq:CurvaturaExtrinsicaij}
\end{eqnarray}
where

\begin{eqnarray}
\dot{t}=\sqrt{\frac{N^{2}+\mc{B}\dot{a}^{2}}{\mc{A}}},
\end{eqnarray}
we found this by taking into consideration
\begin{eqnarray}
U_{M}U^{M}=-1, \quad n_{M}U^{M}=0, \quad n_{M}n^{M}=1,
\end{eqnarray}
where $U^{A}=\left(\frac{\dot{t}}{N},\frac{\dot{a}}{N},0,...,0\right)$ and $n_{A}$ is the normal vector to the brane, where
\begin{eqnarray}
n^{\pm}_{0}=\pm \sqrt{\mc{AB}}\frac{\dot{a}}{N}, \quad n^{\pm}_{a}=\mp \sqrt{\mc{AB}}\frac{\dot{t}}{N}.
\end{eqnarray}
We consider that the unperturbed energy-momentum tensor must have the form of a perfect fluid, therefore its components are those found in \eqref{eq:ComponentesTensorFluidoPerfecto}. Then the source tensor is
\begin{eqnarray}
S_{\tau\tau}=-\left(\frac{D-1}{D}\rho+P\right)N^{2}, \quad S_{i\tau}=0, \quad S_{ij}=-\frac{1}{D}\rho a^{2}\tilde{\gamma}_{ij}, \nonumber \\
\end{eqnarray}
and we find that the density and pressure are
\begin{eqnarray}
\rho =-\frac{D}{2a^{2}\kappa }\left(\sqrt{\mc{B}^{-1}_{+}+\frac{\dot{a}^{2}}{N^{2}}}+\sqrt{\mc{B}^{-1}_{-}+\frac{\dot{a}^{2}}{N^{2}}}\right)\partial_{a}\mc{C},
\end{eqnarray}

\begin{eqnarray}
P=&&\frac{1}{2N^{2}\kappa }\left[\frac{1}{\dot{t}_{+}\sqrt{\mc{A}_{+}\mc{B}_{+}}}\left(\dot{t}^{2}_{+}\partial_{a}\mc{A}_{+}+\dot{a}^{2}\partial_{a}\mc{B}_{+}+2\mc{B}_{+}\left(\ddot{a}-\frac{\dot{a}\dot{N}}{N^{2}}\right)\right.\right.\nonumber\\
&&+\left.\left.\frac{\dot{a}^{2}\dot{t}^{2}}{N^{2}}\left(1-N\right)\left(\mc{B}_{+}\partial_{a}\mc{A}_{+}+\mc{A}_{+}\partial_{a}\mc{B}_{+}\right)\right)+\left(+\rightarrow-\right) \right] \nonumber\\
&&+\frac{D-1}{2a^{2}\kappa }\left(\sqrt{\mc{B}^{-1}_{+}+\frac{\dot{a}^{2}}{N^{2}}}+\sqrt{\mc{B}^{-1}_{-}+\frac{\dot{a}^{2}}{N^{2}}}\right)\partial_{a}\mc{C}.
\end{eqnarray}
The bulk metric perturbation is $h^{\pm}_{AB}$ and the brane metric perturbation is $\eta_{\mu\nu}$. Thus, the source perturbation tensor is
\begin{eqnarray}
\delta S_{\mu\nu}=-\delta T_{\mu\nu}+\frac{1}{D}\left(\gamma_{\mu\nu}\delta T^{\lambda}_{\ \lambda}+T^{\lambda}_{\ \lambda}\eta_{\mu\nu}\right),
\end{eqnarray}
and the extrinsic curvature perturbation is
\begin{eqnarray}
\delta K_{\mu\nu}=-\xi^{A}_{\mu}\partial_{\nu}\delta n_{A}+\xi^{A}_{\mu}\xi^{B}_{\nu}\left(\Gamma^{C}_{AB}\delta n_{C}+n_{C}\delta\Gamma^{C}_{AB} \right).
\end{eqnarray}
Hence, the perturbed part of the equation of motion is
\begin{eqnarray}
\left\langle \delta K_{\mu\nu} \right\rangle=\kappa  \delta S_{\mu\nu},
\end{eqnarray}
or explicitly
\begin{eqnarray}
\left\langle -\xi^{A}_{\mu}\partial_{\nu}\delta n_{A}+\xi^{A}_{\mu}\xi^{B}_{\nu}\left(\Gamma^{C}_{AB}\delta n_{C}+n_{C}\delta\Gamma^{C}_{AB} \right) \right\rangle=\kappa  \left[-\delta T_{\mu\nu}+\frac{1}{D}\left(\gamma_{\mu\nu}\delta T^{\lambda}_{\ \lambda}+T^{\lambda}_{\ \lambda}\eta_{\mu\nu}\right)\right]. \nonumber \\ 
\end{eqnarray}
The perturbations of the Christoffel symbols are
\begin{eqnarray}
\delta\Gamma^{0}_{00}&=&\frac{1}{2\mc{A}}\left(\frac{1}{\mc{B}}h_{0a}\partial_{a}\mc{A} -\partial_{0}h_{00}\right),\nonumber\\
\delta\Gamma^{0}_{0a}&=&\frac{1}{2\mc{A}}\left(\frac{1}{\mc{A}}h_{00}\partial_{a}\mc{A}-\partial_{a}h_{00}\right),\nonumber\\
\delta\Gamma^{0}_{0i}&=&-\frac{1}{2\mc{A}}\partial_{i}h_{00}, \nonumber\\
\delta\Gamma^{0}_{aa}&=&\frac{1}{2\mc{A}}\left(\frac{1}{\mc{B}}h_{0a}\partial_{a} \mc{B}-2\partial_{a}h_{0a}+\partial_{0}h_{aa}\right),\nonumber\\
\delta\Gamma^{0}_{ij}&=&\frac{1}{2\mc{A}}\left(-\frac{1}{\mc{B}}h_{0a}\tilde{g}_{ij}\partial_{a}\, \mc{C} -\nabla_{j}h_{0i}-\nabla_{i}h_{0j}+\partial_{0}h_{ij}\right), \nonumber\\
\delta\Gamma^{0}_{ai}&=&\frac{1}{2\mc{A}}\left(\frac{1}{\mc{C}}h_{0i}\partial_{a} \mc{C} -\partial_{i}h_{0a}-\partial_{a}h_{0i}+\partial_{0}h_{ai}\right),
\end{eqnarray}

\begin{eqnarray}
\delta\Gamma^{a}_{aa}&=&\frac{1}{2\mc{B}}\left(-\frac{1}{\mc{B}}h_{aa}\partial_{a}\mc{B}+\partial_{a}h_{aa}\right),  \nonumber\\ 
\delta\Gamma^{a}_{0a}&=&\frac{1}{2\mc{B}}\left(-\frac{1}{\mc{A}}h_{0a}\partial_{a} \mc{A}+\partial_{0}h_{aa}\right),\nonumber\\
\delta\Gamma^{a}_{ai}&=&\frac{1}{2\mc{B}}\left(-\frac{1}{\mc{C}}h_{ai}\partial_{a} \mc{C} +\partial_{i}h_{aa}\right),\nonumber\\
\delta\Gamma^{a}_{00}&=&\frac{1}{2\mc{B}}\left(-\frac{1}{\mc{B}}h_{aa}\partial_{a}\, \mc{A}+2\partial_{0}h_{0a}-\partial_{a}h_{00}\right),\nonumber\\
\delta\Gamma^{a}_{ij}&=&\frac{1}{2\mc{B}}\left(\frac{1}{\mc{B}}h_{aa}\tilde{g}_{ij}\partial_{a}\, \mc{C}+\nabla_{j}h_{ai}+\nabla_{i}h_{aj}-\partial_{a}h_{ij}\right),\nonumber\\
\delta\Gamma^{a}_{0i}&=&\frac{1}{2\mc{B}}\left(\partial_{i}h_{0a}+\partial_{0}h_{ai}-\partial_{a}h_{0i}\right),
\end{eqnarray}
   
\begin{eqnarray}
\delta\Gamma^{i}_{jk}&=&\frac{1}{2 \mc{C}}\tilde{g}^{il}\left(\frac{1}{ \mc{B}}h_{al}\tilde{g}_{jk}\partial_{a}  \mc{C}+\nabla_{k}h_{lj}+\nabla_{j}h_{lk}-\nabla_{l}h_{jk}\right),\nonumber\\
\delta\Gamma^{i}_{0j}&=&\frac{1}{2 \mc{C}}\tilde{g}^{il}\left(\nabla_{j}h_{0l}+\partial_{0}h_{lj}-\nabla_{l}h_{0j}\right),\nonumber\\
\delta\Gamma^{i}_{aj}&=&\frac{1}{2 \mc{C}}\tilde{g}^{il}\left(-\frac{1}{ \mc{C}}h_{lj}\partial_{a}  \mc{C} +\nabla_{j}h_{al}+\partial_{a}h_{lj}-\nabla_{l}h_{aj}\right),\nonumber\\
\delta\Gamma^{i}_{00}&=&\frac{1}{2 \mc{C}}\tilde{g}^{il}\left(-\frac{1}{ \mc{B}}h_{al}\partial_{a}\,  \mc{A}  +2\partial_{0}h_{0l}-\partial_{l}h_{00}\right), \nonumber\\
\delta\Gamma^{i}_{aa}&=&\frac{1}{2 \mc{C}}\tilde{g}^{il}\left(-\frac{1}{ \mc{B}}h_{al}\partial_{a}  \mc{B}+2\partial_{a}h_{la}-\partial_{l}h_{aa}\right), \nonumber\\
\delta\Gamma^{i}_{0a}&=&\frac{1}{2 \mc{C}}\tilde{g}^{il}\left(-\frac{1}{ \mc{A}}h_{0l}\partial_{a}  \mc{A}+\partial_{a}h_{0l}+\partial_{0}h_{al}-\partial_{l}h_{0a}\right),
\end{eqnarray}
It is possible to calculate the perturbations of extrinsic curvature
\begin{eqnarray}
\delta K^{\pm}_{\tau\tau}&=&\pm\left[-\dot{t}\delta\dot{n}_{0}-\dot{a}\delta\dot{n}_{a}+\frac{1}{2\mc{AB}}\left(2\mc{B}\dot{a}\dot{t}\partial_{a}\mc{A}\delta n_{0}+\mc{A}\left(\dot{t}^{2}\partial_{a}\mc{A}+\dot{a}^{2}\partial_{a}\mc{B}\right)\delta n_{a}\right)\right.\nonumber\\
&&+\frac{\sqrt{\mc{AB}}}{2N\mc{A}^{2}\mc{B}^{2}}\left( 2\mc{B}^{2}\dot{a}^{2}\dot{t}\partial_{a}\mc{A}h_{00}+\mc{A}^{2}\dot{t}\left(\dot{t}^{2}\partial_{a}\mc{A}+\dot{a}^{2}\partial_{a}\mc{B}\right)h_{aa}+\mc{AB}\dot{a}\left(3\dot{t}^{2}\partial_{a}\mc{A}+\dot{a}^{2}\partial_{a}\mc{B}\right)h_{0a}\right)\nonumber\\
&&-\frac{1}{2N\sqrt{\mc{AB}}}\left(\mc{B}\dot{a}\dot{t}^{2}\partial_{0}h_{00}+2\mc{A}\dot{t}^{3}\partial_{0}h_{0a}+\left(2\mc{A}\dot{t}^{2}\dot{a}-\mc{B}\dot{a}^{3}\right)\partial_{0}h_{aa}    \right)\nonumber\\
&&\left.-\frac{1}{2N\sqrt{\mc{AB}}}\left(2\mc{B}\dot{a}^{3}\partial_{a}h_{0a}+\mc{A}\dot{t}\dot{a}^{2}\partial_{a}h_{aa}+\left(2\mc{B}\dot{a}^{2}\dot{t}-\mc{A}\dot{t}^{3}\right)\partial_{a}h_{00}\right)\right],
\end{eqnarray}

\begin{eqnarray}
\delta K^{\pm}_{\tau i}&=&\pm\left[-\delta \dot{n}_{i}+\frac{\dot{a}\partial_{a}\mc{C} }{2\mc{C}}\delta n_{i}+\frac{1}{2N\sqrt{\mc{AB}}}\left(\frac{\dot{a}\partial_{a}\mc{C}}{\mc{C}}\left(\mc{B}\dot{a}h_{0i}+\mc{A}\dot{t}h_{ai}\right)+N^{2}\left(\partial_{a}h_{0i}-\partial_{0}h_{ai}\right)\right.\right. \nonumber\\
&&\left.-\left.\frac{}{}\left(\mc{B}\dot{a}\dot{t}\partial_{i}h_{00}+\left(\mc{A}\dot{t}^{2}+\mc{B}\dot{a}^{2}\right)\partial_{i}h_{0a}+\mc{A}\dot{a}\dot{t}\partial_{i}h_{aa}\right)\right)\right],
\end{eqnarray}

\begin{eqnarray}
\delta K^{\pm}_{ij}&=&\pm\left[\frac{1}{2N\sqrt{\mc{AB}}}\left(-\frac{\partial_{a} \mc{C}}{\mc{B}}\tilde{g}_{ij}\left(\mc{B}\dot{a}h_{0a}+\mc{A}\dot{t}h_{aa}\right)-\mc{B}\dot{a}\left(\nabla_{j}h_{0 i}+\nabla_{i}h_{0 j}-\partial_{0}h_{ij}\right)\right.\right.\nonumber\\
&&-\left.\mc{A}\dot{t}\left(\nabla_{j}h_{a i}+\nabla_{i}h_{a j}-\partial_{a}h_{ij}\right)\biggr)-\frac{\partial_{a} \mc{C}}{2 \mc{B}}\tilde{g}_{ij}\delta n_{a}-\nabla_{i}\delta n_{j}\right].
\end{eqnarray}
The relationship between brane perturbations and bulk perturbations are $\eta_{\mu\nu}=\xi^{M}_{\mu}\xi^{N}_{\nu}h_{MN}$ and it is the same relationship for all tensors involved, thus
\begin{eqnarray}
\eta_{\tau\tau}&=&\dot{t}^{2}h_{00}+2\dot{a}\dot{t}h_{0a}+\dot{a}^{2}h_{aa} \nonumber\\
\eta_{\tau i}&=&\dot{t}h_{0i}+\dot{a}h_{ai} \nonumber\\
\eta_{ij}&=&h_{ij},
\end{eqnarray}
The perturbations of the equations of Einstein for a perfect fluid takes the form
{\normalsize
\begin{eqnarray}
&&\left\langle \delta K_{\tau\tau} \right\rangle=\kappa \left(-\delta T_{\tau\tau}+\frac{1}{D}\left(-a^{2}\delta T^{\lambda}_{\lambda}+T^{\lambda}_{\lambda}\eta_{\tau\tau}\right)\right)\nonumber\\
&&=\kappa \left(\frac{1-D}{D}a^{2}\delta\rho-a^{2}\delta P-\frac{a^{2}}{D}\tilde{\gamma}^{ml}\nabla_{m}\nabla_{l}\pi^{S}+\left(\frac{D-1}{D}\rho+P\right)\left(\dot{t}^{2}h_{00}+2\dot{a}\dot{t}h_{0a}+\dot{a}^{2}h_{aa}\right)\right),\nonumber\\
\label{eq:EcEinsteinScalarKExt}
\end{eqnarray}

\begin{eqnarray}
\left\langle \delta K_{\tau i} \right\rangle&=&\kappa \left(-\delta T_{\tau i}+\frac{1}{D}T^{\lambda}_{\lambda}\eta_{\tau i}\right)=\kappa \left(a\left(\rho+P\right)\delta u_{i}-\frac{1}{D}\rho\left(\dot{t}h_{0i}+\dot{a}h_{ai}\right)\right),
\label{eq:EcEinsteinVectorKExt}
\end{eqnarray}

\begin{eqnarray}
&&\left\langle \delta K_{ij} \right\rangle=\kappa \left(-\delta T_{ij}+\frac{1}{D}\left(a^{2}\tilde{\gamma}_{ij}\delta T^{\lambda}_{\lambda}+T^{\lambda}_{\lambda}\eta_{ij}\right)\right)\nonumber\\
&&=\kappa \left(-\frac{a^{2}}{D}\tilde{\gamma}_{ij}\delta\rho+\frac{a^{2}}{D}\tilde{\gamma}_{ij}\tilde{\gamma}^{ml}\nabla_{m}\nabla_{l}\pi^{S}-a^{2}\left(\nabla_{i}\nabla_{j}\pi^{S}+\nabla_{i}\pi^{V}_{j}+\nabla_{j}\pi^{V}_{i}+\pi^{T}_{ij} \right)-\frac{1}{D}\rho h_{ij}\right).\nonumber\\
\label{eq:EcEinsteinTensorKExt}
\end{eqnarray}
}
Perturbation of the bulk metric can be decomposed into scalars, divergenceless vectors and divergenceless
traceless symmetric tensors,
\begin{eqnarray}
h_{00}&=&-\mc{A}W^{B}, \quad h_{0a}=\sqrt{\mc{AB}}V^{B}, \quad h_{aa}=\mc{B}U^{B}, \nonumber \\
h_{0i}&=&\sqrt{\mc{AC}}\left(\nabla_{i}P^{B}+Q^{B}_{i}\right), \quad h_{ai}=\sqrt{\mc{BC}}\left(\nabla_{i}R^{B}+S^{B}_{i}\right), \nonumber \\
h_{ij}&=&\mc{C}\left(A^{B}\tilde{g}_{ij}+\nabla_{i}\nabla_{j}B^{B}+\nabla_{i}C^{B}_{j}+\nabla_{j}C^{B}_{i}+D^{B}_{ij}\right),
\label{eq:DescomposicionExt}
\end{eqnarray}
where same conditions to those seen in (\ref{eq:RotarDiverSimet}) are complied, also similarly as $\delta u_{i}$ we have that $\delta n^{\pm}_{i}=\nabla_{i}\delta n^{\pm}+\delta n^{V\pm}_{i}$. Perturbations of the energy-momentum tensor on the brane, as well as their respective conservation conditions, have the form that was developed in Sec. \ref{sec:PertCosmoStandar}.
\newline

Hereafter we show the scalar, vector and tensor modes of Eqs. \eqref{eq:EcEinsteinScalarKExt}-\eqref{eq:EcEinsteinTensorKExt}.

%%%%%%%%%%%%%%%%%%%%%%%%%%%%%%%%%
\subsection{Energy Flux} \label{sec:EF}
%%%%%%%%%%%%%%%%%%%%%%%%%%%%%%%%%

The equations seen in this section generally indicate how when the matter content inside the brane is altered, this implies an alteration in its normal vectors, having repercussions on the bulk. One of the most important repercussions when talking about extra dimensions is knowing if there is an energy flow, the $(\tau, i)$-component of the energy-momentum tensor indicates the energy flow in the $i$-spatial direction, this occurs internally in the brane, However, the equation \eqref{eq:EcEinsteinVector} shows us how the internal geometry of the brane is altered due to this internal flux, and the equation \eqref{eq:EcEinsteinVectorKExt} shows us how the external geometry of the brane is altered due to that same flow. On the other hand, the $(0, a)$-component of the energy-momentum tensor tells us the flow of energy in the bulk in the direction of the $a$-extra dimension \cite{Dorca:Bruck2001}. To find this component, we must take into account that in this paper, we consider that the matter is contained exclusively in the brane and the rest of the bulk is empty, however not necessarily the $(0, a)$-component of the energy-momentum tensor is null, through the transformation $T_{\mu\nu}=\xi^{M}_{\mu}\xi^{N}_{\nu}T_{MN}$, we find the component $T_{0a}$, which is obtained from

\begin{eqnarray}
T_{MN}=Pg_{MN}+\left( P+\rho\right)U_{M}U_{N}.
\end{eqnarray}
For the case of perturbations, by performing the same transformation to the equation \eqref{eq:EcEinsteinScalarKExt}, we can extract the part corresponding to the $(0, a)$-component, and we have $\left\langle \delta K_{0a} \right\rangle = \delta T_{0a}$ or explicitly

\begin{eqnarray}
&&\left\langle -\frac{1}{2}\left(\partial_{a}\delta n_{0}+\partial_{0}\delta n_{a}\right)+\frac{1}{2\mc{AB}}\left(\mc{B}\partial_{a}\mc{A}\delta n_{0}+\mc{A}\partial_{0}\mc{B}\delta n_{a}\right)\right.\nonumber\\
&&\left.+\frac{\partial_{a}\mc{A}}{2\mc{A}^{2}\mc{B}^{2}}\left(\mc{B}^{2}n_{0}h_{00}-\mc{AB}n_{a}h_{0a}\right)-\frac{1}{2\mc{AB}}\left(n_{0}\mc{B}\partial_{a}h_{00}-n_{a}\mc{A}\partial_{0}h_{aa}\right)\right\rangle\nonumber\\
&&=-\left( \rho+P \right)\left(U_{0}\delta U_{a}+U_{a}\delta U_{0}\right)-\left( \delta\rho+\delta P \right)U_{0}U_{a}-\frac{\rho}{D}h_{0a},
\end{eqnarray}
which indicates the dynamics of the energy flow in the $a$-direction, due to the perturbations of the brane.

%%%%%%%%%%%%%%%%%%%%%%%%%%%%%%%%%%%%%%%%%%%
\subsection{Scalar Modes}
%%%%%%%%%%%%%%%%%%%%%%%%%%%%%%%%%%%%%%%%%%%

The part of Eq. \eqref{eq:EcEinsteinTensorKExt} proportional to $\tilde{g}_{jk}$ gives

\begin{eqnarray}
&&\left\langle\frac{1}{2N\sqrt{\mc{AB}}}\left(-\frac{\partial_{a} \mc{C}}{\mc{B}}\left(\mc{B}\sqrt{\mc{AB}}\dot{a}V^{B}+\mc{A}\mc{B}\dot{t}U^{B}\right)+\mc{BC}\dot{a}\partial_{0}A^{B}+\mc{A}\dot{t}\left(A^{B}\partial_{a}\mc{C}+\mc{C}\partial_{a}A^{B}\right)\right)\right.\nonumber\\
&&-\left.\frac{\partial_{a} \mc{C}}{2 \mc{B}}\delta n_{a}\right\rangle=-\kappa \frac{a^{2}}{D}\left(\delta\rho+\rho A^{B}-\tilde{\gamma}^{lm}\nabla_{m}\nabla_{l}\pi^{S}\right).
\label{eq:MScalar1KExt}
\end{eqnarray}
The part of Eq. \eqref{eq:EcEinsteinTensorKExt} of the form $\nabla_{j}\nabla_{k}S$ (where $S$ is any scalar) gives
\begin{eqnarray}
&&\nabla_{j}\nabla_{k}\left\{\left\langle\frac{1}{2N\sqrt{\mc{AB}}}\left(-\mc{B}\dot{a}\left(2\sqrt{\mc{AC}}P^{B}-\mc{C}\partial_{0}B^{B}\right)-\mc{A}\dot{t}\left(2\sqrt{\mc{BC}}R^{B}-\mc{C}\partial_{a}B^{B}-B^{B}\partial_{a}\mc{C}\right)\right)\right.\right.\nonumber\\
&&\left.-\delta n\biggr\rangle
+\kappa a^{2}\left(\pi^{S}+\frac{\rho }{D}B^{B}\right)\right\}=0.
\label{eq:MScalar2KExt}
\end{eqnarray}
The part of Eq. \eqref{eq:EcEinsteinVectorKExt} of the form $\nabla_{j}S$ (where $S$ is any scalar) gives
\begin{eqnarray}
& &\left\langle-\nabla_{i}\delta \dot{n}+\frac{\dot{a}\partial_{a}\mc{C} }{2\mc{C}}\nabla_{i}\delta n+\frac{1}{2N\sqrt{\mc{AB}}}\left(\frac{\dot{a}\partial_{a}\mc{C}}{\mc{C}}\left(\mc{B}\sqrt{\mc{AC}}\dot{a}\nabla_{i}P^{B}+\mc{A}\sqrt{\mc{BC}}\dot{t}\nabla_{i}R^{B}\right)\right.\right. \nonumber\\
&&+N^{2}\left(\nabla_{i}P^{B}\partial_{a}\sqrt{\mc{AC}}+\sqrt{\mc{AC}}\nabla_{i}\partial_{a}P^{B}-\sqrt{\mc{BC}}\nabla_{i}\partial_{0}R^{B}\right) \nonumber\\
&&\left.\left.-\left(-\mc{AB}\dot{a}\dot{t}\nabla_{i}W^{B}+\sqrt{\mc{AB}}\left(\mc{A}\dot{t}^{2}+\mc{B}\dot{a}^{2}\right)\nabla_{i}V^{B}+\mc{AB}\dot{a}\dot{t}\nabla_{i}U^{B}\right)\right)\right\rangle \nonumber\\
&&=-\kappa \left(\frac{\sqrt{\mc{C}}\rho }{D}\nabla_{i}\left(\dot{t}\sqrt{\mc{A}}P^{B}+\dot{a}\sqrt{\mc{B}}R^{B}\right)-a\left(P +\rho \right)\nabla_{i}\delta u\right).
\label{eq:MScalar3KExt}
\end{eqnarray}
Eq. \eqref{eq:EcEinsteinScalarKExt} gives
\begin{eqnarray}
&&\left\langle-\dot{t}\delta\dot{n}_{0}-\dot{a}\delta\dot{n}_{a}+\frac{1}{2\mc{AB}}\left(2\mc{B}\dot{a}\dot{t}\partial_{a}\mc{A}\delta n_{0}+\mc{A}\left(\dot{t}^{2}\partial_{a}\mc{A}+\dot{a}^{2}\partial_{a}\mc{B}\right)\delta n_{a}\right)\right.\nonumber\\
&&+\frac{\sqrt{\mc{AB}}}{2N\mc{A}^2\mc{B}^2}\left(-2\mc{B}^2\dot{a}^2\dot{t}\mc{A}W^B\partial_a\mc{A}+\mc{A}^2\mc{B}\dot{t}\left(\dot{t}^2\partial_a\mc{A}+\dot{a}^2\partial_a\mc{B}\right)U^{B}\right.\nonumber\\
&&+\left.\mc{AB}\sqrt{\mc{AB}}\dot{a}\left(3\dot{t}^{2}\partial_{a}\mc{A}+\dot{a}^{2}\partial_{a}\mc{B}\right)V^{B}\right)\nonumber\\
&&-\frac{1}{2N\sqrt{\mc{AB}}}\left(-\mc{A}\mc{B}\dot{a}\dot{t}^{2}\partial_{0}W^{B}+2\mc{A}\sqrt{\mc{AB}}\dot{t}^{3}\partial_{0}V^{B}+\mc{B}\left(2\mc{A}\dot{t}^{2}\dot{a}-\mc{B}\dot{a}^{3}\right)\partial_{0}U^{B}\right)\nonumber\\
&&-\frac{1}{2N\sqrt{\mc{AB}}}\left(2\mc{B}\dot{a}^{3}\left(\sqrt{\mc{AB}}\partial_{a}V^{B}+V^{B}\partial_{a}\sqrt{\mc{AB}}\right)+\mc{A}\dot{t}\dot{a}^{2}\left(\mc{B}\partial_{a}U^{B}+U^{B}\partial_{a}\mc{B}\right)\right.\nonumber\\
&&
-\left(2\mc{B}\dot{a}^{2}\dot{t}-\mc{A}\dot{t}^{3}\right)\left(\mc{A}\partial_{a}W^{B}+W^{B}\partial_{a}\mc{A}\right)\biggr)\bigg\rangle\nonumber\\
&&=\kappa \left(\frac{1-D}{D}a^{2}\delta\rho-a^{2}\delta P-\frac{a^{2}}{D}\tilde{\gamma}^{ml}\nabla_{m}\nabla_{l}\pi^{S}+\left(\frac{D-1}{D}\rho+P\right)\left(-\dot{t}^{2}\mc{A}W^{B}\right.\right.\nonumber \\ &&+\left.\left.2\dot{a}\dot{t}\sqrt{\mc{AB}}V^{B}+\dot{a}^{2}\mc{B}U^{B}\right)\right).\nonumber\\
\label{eq:MScalar4KExt}
\end{eqnarray}

%%%%%%%%%%%%%%%%%%%%%%%%%%%%%%%%%%%%%%%%%%%
\subsection{Vector Modes}
%%%%%%%%%%%%%%%%%%%%%%%%%%%%%%%%%%%%%%%%%%%

The part of Eq. \eqref{eq:EcEinsteinTensorKExt} of the form $\nabla_{k}V_{j}$ (where $V_{j}$ is a divergenceless vector) gives
\begin{eqnarray}
&&\left\langle\frac{1}{2N\sqrt{\mc{AB}}}\left(-\mc{B}\dot{a}\left(\sqrt{\mc{AC}}\nabla_{i}Q^{B}_{j}-\mc{C}\partial_{0}\nabla_{i}C^{B}_{j}\right)-\mc{A}\dot{t}\left(\sqrt{\mc{BC}}\nabla_{i}S^{B}_{j}-\left(\mc{C}\nabla_{i}\partial_{a}C^{B}_{j}+\nabla_{i}C^{B}_{j}\partial_{a}\mc{C}\right)\right)\right)\right.\nonumber\\
&&\left.-\nabla_{i}\delta n^{V}_{j}\frac{}{}\right\rangle=-\kappa \left(a^{2}\nabla_{i}\pi^{V}_{j}+\frac{a^{2}\rho}{D}\nabla_{i}C_{j}\right),
\end{eqnarray}
while the part of Eq. (\ref{eq:EcEinsteinVectorKExt}) of the form $V_{j}$ (where $V_{j}$ is any divergenceless vector) gives
\begin{eqnarray}
&&\left\langle-\delta \dot{n}^{V}_{i}+\frac{\dot{a}\partial_{a}\mc{C} }{2\mc{C}}\delta n^{V}_{i}+\frac{1}{2N\sqrt{\mc{AB}}}\left(\frac{\dot{a}\partial_{a}\mc{C}}{\sqrt{\mc{C}}}\left(\mc{B}\sqrt{\mc{A}}\dot{a}Q^{B}_{i}+\mc{A}\sqrt{\mc{B}}\dot{t}S^{B}_{i}\right)\right.\right.\nonumber\\
&&\left.\left.+N^{2}\left(\left(\sqrt{\mc{AC}}\partial_{a}Q^{B}_{i}+Q^{B}_{i}\partial_{a}\sqrt{\mc{AC}}\right)-\sqrt{\mc{BC}}\partial_{0}S^{B}_{i}\right)\right)\right\rangle=\kappa \left(a\left(\rho+P\right)\delta u^{V}_{i}-\frac{1}{D}\rho aN G_{i}\right).\nonumber\\
\end{eqnarray}

%%%%%%%%%%%%%%%%%%%%%%%%%%%%%%%%%%%%%%%%%%%
\subsection{Tensor Modes}
%%%%%%%%%%%%%%%%%%%%%%%%%%%%%%%%%%%%%%%%%%%

The part of Eq. \eqref{eq:EcEinsteinTensorKExt} of the form of a traceless and divergenceless tensors is
\begin{eqnarray}
\left\langle\frac{1}{2N\sqrt{\mc{AB}}}\left[\mc{B}\dot{a}a^{2}\partial_{0}D_{ij}+\mc{A}\dot{t}\left(a^{2}\partial_{a}D_{ij}+2aD_{ij}\right)\right]\right\rangle=-\kappa a^{2}\left(\pi^{T}_{ij}+\frac{\rho}{D}D_{ij}\right),
\label{eq:ModoTensorBrana}
\end{eqnarray}
the conditions of conservation of the current are internal properties of the brane, therefore they are the same as in previous equations.

%%%%%%%%%%%%%%%%%%%%%%%%%%%%%%%%%%%%%%%%%%%%%%%%
\section{Gauges in Perturbations of branes}  \label{sec:GaugesBrane}
%%%%%%%%%%%%%%%%%%%%%%%%%%%%%%%%%%%%%%%%%%%%%%%%

We have been studying up to this point the perturbations of the Bulk, its decompositions, its relationship with the intrinsic perturbations of the brane, the latter through covariant transformations, from this we have derived equations of motion, equations that can be complicated to solve. However, some perturbations may lack physical meaning, it is of great interest to observe if by discarding such perturbations, the equations of motion remain invariant, in addition to possibly being “easier” to solve.

For this, a study with Gauge transformations must be carried out. In the section \ref{sec:Gauges} the gauge transformations and choices for Cosmology in D+1 dimensions have been carried out, a topic that has already been discussed in the literature for the case $D=3$. Now we want to apply the respective gauge transformations to the bulk perturbations in order to keep the perturbation equations invariant. Our treatment will be similar to that shown in the \ref{sec:Gauges} section.

Here, we consider a space-time coordinate transformation para el Bulk as \cite{weinbergcosmology}
\begin{eqnarray}
x^{A}\rightarrow x'^{A}=x^{A}+e^{A}\left(x\right).
\label{eq:TransformacionCoordenadaBulk}
\end{eqnarray}
Equations \eqref{eq:Deltah1} and \eqref{eq:DeltaZ1} can be applied to bulk, since they are carried out for arbitrary metrics and second order tensors in general, thus we obtain the transformations $\Delta h_{AB}$ for our bulk in particular

\begin{eqnarray}
&\Delta h_{00}=-2\partial_{0} e_{0}+\frac{1}{\mc{B}}e_{a}\partial_{a}\mc{A}, \quad \Delta h_{0a}=-\partial_{0} e_{a}-\partial_{a} e_{0}+\frac{1}{\mc{A}}e_{0}\partial_{a}\mc{A},&\nonumber \\[5pt] &\Delta h_{aa}=-2\partial_{a} e_{a}+\frac{1}{\mc{B}}e_{a}\partial_{a}\mc{B},&\nonumber \\[5pt]
&\Delta h_{0i}=-\partial_{0} e_{i}-\nabla_{i}e_{0}, \quad \Delta h_{ai}=-\partial_{a} e_{i}-\nabla_{i}e_{a}+\frac{1}{\mc{C}}e_{i}\partial_{a}\mc{C},&\nonumber \\[5pt]
&\Delta h_{ij}=-\nabla_{j}e_{i}-\nabla_{i}e_{j}-\frac{1}{\mc{B}}\tilde{g}_{ij}e_{a}\partial_{a}\mc{C}.&
\label{eq:TransformacioneshExt}
\end{eqnarray}
and the gauge transformations for a second order tensor of our bulk are

\begin{eqnarray}
&\Delta \delta Z_{00}=-\zeta\Delta h_{00}+e_{0}\partial_{0}\zeta-\frac{\mc{A}}{\mc{B}}e_{a}\partial_{a}\zeta, \quad \Delta \delta Z_{0a}=\zeta\partial_{a}e_{0}-\frac{1}{\mc{A}}e_{0}\zeta\partial_{a}\mc{A}-\chi\partial_{0}e_{a},&\nonumber \\[5pt]
&\Delta \delta Z_{aa}=\chi\Delta h_{aa}-e_{a}\partial_{a}\chi+\frac{\mc{B}}{\mc{A}}e_{0}\partial_{0}\chi,&\nonumber \\[5pt]
&\Delta \delta Z_{0i}=\zeta\partial_{i}e_{0}-\xi\partial_{0}e_{i}, \quad \Delta \delta Z_{ai}=-\chi\partial_{i}e_{a}+\frac{1}{\mc{C}}e_{i}\chi\partial_{a}\mc{C}-\xi\partial_{a}e_{i},&\nonumber \\[5pt]
&\Delta \delta Z_{ij}=\xi\Delta h_{ij}+\mc{C} \tilde{g}_{ij}\left(\frac{1}{\mc{A}}e_{0}\partial_{0}\xi-\frac{1}{\mc{B}}e_{a}\partial_{a}\xi\right),&.
\label{eq:TransformacionesZExt}
\end{eqnarray}
where $Z_{00}=\zeta(t,a)\mc{A}$, $Z_{aa}=\chi(t,a)\mc{B}$, $Z_{ij}=\xi(t,a)\mc{C}\tilde{g}_{ij}$.\\

The covariant transformation $T_{\mu\nu}=\xi^{A}_{\mu}\xi^{B}_{\nu}T_{AB}$ \cite{dodelsoncosmology} between energy-momentum tensors indicates that in particular  $\zeta=-\chi=\rho$ and $\xi=P$. Then, Eq. \eqref{eq:TransformacioneshExt} gives the gauge transformations of the metric perturbation components defined by Eqs. \eqref{eq:DescomposicionExt}\cite{Kodama:Sasak1984}:
\begin{eqnarray}
&\Delta W^{B}=\frac{1}{\mc{A}}\left(2\partial_{0} e_{0}-\frac{1}{\mc{B}}e_{a}\partial_{a}\mc{A} \right), \quad \Delta V^{B}=\frac{1}{\sqrt{\mc{AB}}}\left(-\partial_{0} e_{a}-\partial_{a} e_{0}+\frac{1}{\mc{A}}e_{0}\partial_{a}\mc{A}\right),&\nonumber \\
&\Delta U^{B}=\frac{1}{\mc{B}}\left( -2\partial_{a} e_{a}+\frac{1}{\mc{B}}e_{a}\partial_{a}\mc{B} \right),&\nonumber \\[5pt]
&\Delta P^{B}=-\frac{1}{\sqrt{\mc{AC}}}\left(\partial_{0}e^{S}+e_{0}\right), \quad \Delta Q^{B}_{i}=-\frac{1}{\sqrt{\mc{AC}}}\partial_{0}e^{V}_{i},&\nonumber \\[5pt]
&\Delta R^{B}=\frac{1}{\sqrt{\mc{BC}}}\left(-\partial_{a}e^{S}-e_{a}+\frac{1}{\mc{C}}e^{S}\partial_{a}\mc{C}\right), \quad \Delta S^{B}_{i}=\left(\frac{1}{\sqrt{\mc{BC}}}\partial_{a}e^{V}_{i}+\frac{1}{\mc{C}}e^{V}_{i}\partial_{a}\mc{C}\right),&\nonumber \\[5pt]
&\Delta A^{B}=-\frac{1}{\mc{BC}}e_{a}\partial_{a}\mc{C}, \quad \Delta B^{B}=-\frac{2}{\mc{C}}e^{S}, \quad \Delta C^{B}_{i}=-\frac{1}{\mc{C}}e_{i}^{V}, \quad \Delta D^{B}_{ij}=0,&
\end{eqnarray}
The transformation $\Delta\delta T_{\mu\nu}=\xi^{A}_{\mu}\xi^{B}_{\nu}\Delta\delta T_{AB}$ and \eqref{eq:TransformacionesZExt}, indicates that the gauge transformations for the perturbations to the pressure, energy density, velocity potential and dissipative terms, are the same as those shown in \eqref{eq:TGaugesRhoP} and \eqref{eq:GuageInvariant}.

%%%%%%%%%%%%%%%%%%%%%%%%%%%%%%%%%%%%
\subsection{Newtonian gauge}
%%%%%%%%%%%%%%%%%%%%%%%%%%%%%%%%%%%

The bulk line element for this gauge is \cite{weinbergcosmology,dodelsoncosmology,Ma_1995}
\begin{eqnarray}
ds^{2}= -\mc{A}_{\pm}\left( 1+2\Xi_{\pm} \right)dt^{2}_{\pm}+4\sqrt{\mc{A_{\pm}B_{\pm}}}\Upsilon_{\pm} dtda+\mc{B}_{\pm}\left( 1+2\Theta_{\pm} \right)da^{2}+a^{2}(1-2\Psi)\tilde{g}_{ij}dx^{i}dx^{j}, \nonumber \\
\end{eqnarray}
then $P^{B}$, $R^{B}$ and $B^{B}$ are null. The gravitational field equations \eqref{eq:MScalar1KExt}-\eqref{eq:MScalar4KExt}, takes the form
\begin{eqnarray}
&&\left\langle \frac{1}{N\sqrt{\mc{AB}}}\left(-\frac{\partial_{a} \mc{C}}{\mc{B}}\left(\mc{B}\sqrt{\mc{AB}}\dot{a}\Upsilon+\mc{A}\mc{B}\dot{t}\Theta\right)-\mc{BC}\dot{a}\partial_{0}\Psi-\mc{A}\dot{t}\left(\Psi\partial_{a}\mc{C}+\mc{C}\partial_{a}\Psi\right)\right)-\frac{\partial_{a} \mc{C}}{2 \mc{B}}\delta n_{a}\right\rangle \nonumber\\
&&=-\kappa \frac{a^{2}}{D}\left(\delta\rho-2\rho \Psi-\tilde{\gamma}^{lm}\nabla_{m}\nabla_{l}\pi^{S}\right),
\label{eq:Newton1KExt}
\end{eqnarray}

\begin{eqnarray}
\nabla_{j}\nabla_{k}\left\{-\delta n^{+}+\delta n^{-}+\kappa a^{2}\pi^{S}\right\}=0,
\label{eq:Newton2KExt}
\end{eqnarray}

\begin{eqnarray}
&&\left\langle -\nabla_{i}\delta \dot{n}+\frac{\dot{a}\partial_{a}\mc{C} }{2\mc{C}}\nabla_{i}\delta n-\frac{1}{N\sqrt{\mc{AB}}}\left(-\mc{AB}\dot{a}\dot{t}\nabla_{i}\Xi+\sqrt{\mc{AB}}\left(\mc{A}\dot{t}^{2}+\mc{B}\dot{a}^{2}\right)\nabla_{i}\Upsilon+\mc{AB}\dot{a}\dot{t}\nabla_{i}\Theta\right)\right\rangle \nonumber \\
&&=\kappa a\left(P +\rho \right)\nabla_{i}\delta u, \nonumber\\
\label{eq:Newton3KExt}
\end{eqnarray}

\begin{eqnarray}
&&\left\langle-\dot{t}\delta\dot{n}_{0}-\dot{a}\delta\dot{n}_{a}+\frac{1}{2\mc{AB}}\left(2\mc{B}\dot{a}\dot{t}(\partial_{a}\mc{A})\delta n_{0}+\mc{A}\left(\dot{t}^{2}\partial_{a}\mc{A}+\dot{a}^{2}\partial_{a}\mc{B}\right)\delta n_{a}\right)\right.\nonumber\\
&&+\frac{\sqrt{\mc{AB}}}{N\mc{A}^{2}\mc{B}^{2}}\left(-2\mc{B}^{2}\dot{a}^{2}\dot{t}\mc{A}\Xi\partial_{a}\mc{A}+\mc{A}^{2}\mc{B}\dot{t}\left(\dot{t}^{2}\partial_{a}\mc{A}+\dot{a}^{2}\partial_{a}\mc{B}\right)\Theta+\mc{AB}\sqrt{\mc{AB}}\dot{a}\left(3\dot{t}^{2}\partial_{a}\mc{A}+\dot{a}^{2}\partial_{a}\mc{B}\right)\Upsilon\right)\nonumber\\
&&-\frac{1}{N\sqrt{\mc{AB}}}\left(-\mc{A}\mc{B}\dot{a}\dot{t}^{2}\partial_{0}\Xi+2\mc{A}\sqrt{\mc{AB}}\dot{t}^{3}\partial_{0}\Upsilon+\mc{B}\left(2\mc{A}\dot{t}^{2}\dot{a}-\mc{B}\dot{a}^{3}\right)\partial_{0}\Theta\right)\nonumber\\
&&-\frac{1}{N\sqrt{\mc{AB}}}\left(2\mc{B}\dot{a}^{3}\left(\sqrt{\mc{AB}}\partial_{a}\Upsilon+\Upsilon\partial_{a}\sqrt{\mc{AB}}\right)+\mc{A}\dot{t}\dot{a}^{2}\left(\mc{B}\partial_{a}\Theta+\Theta\partial_{a}\mc{B}\right)\right.\nonumber \\
&&-\left(2\mc{B}\dot{a}^{2}\dot{t}-\mc{A}\dot{t}^{3}\right)\left(\mc{A}\partial_{a}\Xi+\Xi\partial_{a}\mc{A}\right)\Bigr)\biggr\rangle\nonumber\\
&&=\kappa \left(\frac{1-D}{D}a^{2}\delta\rho-a^{2}\delta P-\frac{a^{2}}{D}\tilde{\gamma}^{ml}\nabla_{m}\nabla_{l}\pi^{S}-\left(\frac{D-1}{D}\rho+P\right)2N^{2}\Phi\right).\nonumber\\
\label{eq:Newton4KExt}
\end{eqnarray}
In this last equation, we have used: $\eta_{\tau\tau}=\xi^{A}_{\tau}\xi^{B}_{\tau}h_{AB}$ or $ -N^{2}\Phi=-\dot{t}^{2}\mc{A}\Xi+2\dot{a}\dot{t}\sqrt{\mc{AB}}\Upsilon+\dot{a}^{2}\mc{B}\Theta$.

%%%%%%%%%%%%%%%%%%%%%%%%%%%%%%%%%
\subsection{Synchronous Gauge}
%%%%%%%%%%%%%%%%%%%%%%%%%%%%%%%%%

In the synchronous gauge we have $A^{B}\neq 0$, $B^{B}\neq 0$, all other scalars are null. The line element of such gauge is 
\begin{eqnarray}
ds^{2}=-\mc{A}_{\pm}dt^{2}_{\pm}+\mc{B}_{\pm}da^{2}+a^{2}\left[(1+A^{B})\tilde{g}_{ij}+\nabla_{i}\nabla_{j}B^{B}\right]dx^{i}dx^{j}.
\end{eqnarray}
Hence, the gravitational field equations \eqref{eq:MScalar1KExt}-\eqref{eq:MScalar4KExt}, takes finally the form
\begin{eqnarray}
&&\left\langle\frac{1}{2N\sqrt{\mc{AB}}}\left(\mc{BC}\dot{a}\partial_{0}A^{B}+\mc{A}\dot{t}\left(A^{B}\partial_{a}\mc{C}+\mc{C}\partial_{a}A^{B}\right)\right)-\frac{\partial_{a} \mc{C}}{2 \mc{B}}\delta n_{a}\right\rangle\nonumber\\
&&=-\kappa \frac{a^{2}}{D}\left(\delta\rho+\rho A^{B}-\tilde{\gamma}^{lm}\nabla_{m}\nabla_{l}\pi^{S}\right),
\label{eq:Syncro1KExt}
\end{eqnarray}

\begin{eqnarray}
&&\nabla_{j}\nabla_{k}\left\{\left\langle\frac{1}{2N\sqrt{\mc{AB}}}\left(-\mc{B}\dot{a}\left(-\mc{C}\partial_{0}B^{B}\right)-\mc{A}\dot{t}\left(-\mc{C}\partial_{a}B^{B}-B^{B}\partial_{a}\mc{C}\right)\right)-\delta n\right\rangle\right.\nonumber\\
&&\left.+\kappa a^{2}\left(\pi^{S}+\frac{\rho }{D}B^{B}\right)\right\}=0,
\label{eq:Syncro2KExt}
\end{eqnarray}

\begin{eqnarray}
-\nabla_{i}\delta \dot{n}^{+}+\nabla_{i}\delta \dot{n}^{-}+\frac{\dot{a}\partial_{a}\mc{C} }{2\mc{C}}\left(\nabla_{i}\delta n^{+}-\nabla_{i}\delta n^{-}\right)=\kappa a\left(P +\rho \right)\nabla_{i}\delta u,
\label{eq:Syncro3KExt}
\end{eqnarray}

\begin{eqnarray}
&&\left\langle-\dot{t}\delta\dot{n}_{0}-\dot{a}\delta\dot{n}_{a}+\frac{1}{2\mc{AB}}\left(2\mc{B}\dot{a}\dot{t}(\partial_{a}\mc{A})\delta n_{0}+\mc{A}\left(\dot{t}^{2}\partial_{a}\mc{A}+\dot{a}^{2}\partial_{a}\mc{B}\right)\delta n_{a}\right)\right\rangle\nonumber\\
&&=\kappa \left(\frac{1-D}{D}a^{2}\delta\rho-a^{2}\delta P-\frac{a^{2}}{D}\tilde{\gamma}^{ml}\nabla_{m}\nabla_{l}\pi^{S}\right).\nonumber\\
\label{eq:Syncro4KExt}
\end{eqnarray}

%%%%%%%%%%%%%%%%%%%%%%%%%%%%%%%%%%%%%%
\section{Discussion and Conclusions} \label{Sec:CO}
%%%%%%%%%%%%%%%%%%%%%%%%%%%%%%%%%%%%%%

Here we present a formulation of perturbations in more than the standard 3+1 dimensions, where the background metric is homogeneous and isotropic and consists of an arbitrary sectional curvature where the content of matter is endowed with a cosmological constant, being the geometry of space-time determined by the Ricci tensor as usual. Thus, we develop a set of equations that indicate the evolution of perturbations, where we also propose that the equations of general relativity are fulfilled satisfied with the number of dimensions of space, such equations can be particularized and be very useful for important models such as the case of structure formation of the universe where initial conditions are considered and the content of matter such as photons, baryons and neutrinos, or also the case of generation of gravitational waves, either primordial or in the vicinity of a black hole, both due to tensor perturbations. 

In that same line of thought and looking for a formulation of perturbations in extra dimensions, we present a development of perturbations for branes where we start from evolutionary equations and in where the geometry of space-time is determined by the extrinsic curvature tensor. We have that the bulk where the branes are embedded has a particular distribution of matter, however the homogeneity and isotropy is fulfilled in the internal space of the brane, having in turn an arbitrary sectional curvature. For this case, we also find equations for the evolution of perturbations in branes, and their respective scalar, vector and tensor modes, choosing also the Newtonian conformal and synchronous gauges, these in their version of extra dimensions. 
When we study our equations with the Ricci tensor, we observe the universe as a whole, but when we work with the extrinsic curvature tensor, we observe the universe as part of a whole; in both cases we consider the energy-momentum tensor as a perfect fluid, which already has the structure to also be treated as a scalar fields.

Subsection \ref{sec:EF} deals with energy flows, where we mention that there are internal energy flows in the brane and because of this there can be geometric alterations both internal and external to the brane, in addition that there are external energy flows, all of this when working with extra dimensions and perturbations.

Our work shown in this paper will be very useful for the development of perturbations in the Bulk in a more efficient way, where both the Ricci tensor and the extrinsic curvature will be considered in the same treatment to determine the geometry of the space-time in where we have generalized evolutionary equations for perturbations. Among the possible applications of this generalized model are: i) the generation of gravitational waves in the bulk or in extra dimensions, ii) the formation of structure into the brane due to the interaction with another brane (these may have different matter contents) or even with the same bulk. Furthermore, the linear perturbations shown in this paper has the idea to be applied in novel models with two concentric branes like these studied in \cite{Garc_a_Aspeitia_2010} or in models with variable brane tension \cite{Garcia-Aspeitia:2018fvw} in where the Universe acceleration it is related with the dynamics of extra dimensions. We expect that these models could tackle problems like the amplitude of matter perturbations smoothed over $8h^{-1}$Mpc, $\sigma_8$, where $h$ is the dimensionless Hubble constant. Under the $\Lambda$CDM scenario it is observed that CMB anisotropies are not consistent with weak gravitational lensing measurements \cite{DES:2021wwk,KiDS:2020suj,Heymans_2021}. In fact, the tension is measured through $S_8=\sigma_8\sqrt{\Omega_{0m}/0.3}$, where $\Omega_{0m}$ is the matter density parameter. Thus, the Planck collaboration predicts $S_8=0.834\pm0.016$ while Kilo-Degree Survey (KiDS-1000) \cite{KiDS:2020suj} predicts $S_8=0.766^{+0.020}_{-0.014}$, producing a tension of $3\sigma$ between both results.
Another example where we could apply the equations that we developed is in the perturbations that are generated in the neighborhood of black holes. We can model this by imagining that the black hole is covered by a "Gaussian Surface" that will be analogous to a brane with positive curvature, which could be located on the event horizon, then for this case it is $D=2$ and therefore the brane will have two spatial dimensions (azimuth $\theta$ and zenith $\phi$ coordinates) and one temporal, with the same analogy, the bulk would occupy the place of the background space, being this of 3 spatial dimensions and one temporal, for this model, the radial coordinate $r$ would have the role of extra dimension. In this last example, it is important to note that the model is not the same as the one used in the Arnowitt-Deser-Misner formalism.

We finally mention that all these future studies are only possible under the specific form of the linear perturbations shown in this paper, however these studies will be presented elsewhere.

\section*{Acknowledgments}
We thank the anonymous referee for thoughtful remarks and suggestions. Z.E.M.D appreciates the support of CONACYT and the collaborations of the Escuela Superior de Física y Matemáticas (ESFM) and Instituto Avanzado de Cosmología (IAC). M.A.G-A acknowledges support from c\'atedra Marcos Moshinsky (MM) and Universidad Iberoamericana for the support with the National Research System (SNI) grant. The numerical analysis was carried out by the {\it Numerical Integration for Cosmological Theory and Experiments in High-energy Astrophysics} "Nicte Ha" cluster at IBERO University, acquired through c\'atedra MM support.\\

\bibliography{librero1}

\end{document}